\documentclass[journal]{IEEEtran}
\usepackage[utf8]{inputenc}
\usepackage[T1]{fontenc}
\usepackage{amsmath,amssymb}
\usepackage{booktabs}
\usepackage{multirow}
\usepackage{graphicx}
\usepackage{url}
\usepackage{xcolor}
\usepackage{balance}
\usepackage[hidelinks]{hyperref}
\graphicspath{{figures/}}

\newcommand{\kd}[1]{\texttt{#1}}

\begin{document}

\title{Unknown-Traffic Detection, Calibration and Shortcut Reliance\\in Distilled Encrypted-Traffic Classifiers over One Year}

\author{Mahmoud~Abbasi,~\IEEEmembership{Member,~IEEE}%
\thanks{M. Abbasi was with the University of Salamanca, Salamanca, Spain. He is now with the AIR Institute, Salamanca, Spain (e-mail: mahmoud.abbasi@ieee.org; ORCID 0000-0002-1886-8284).}%
\thanks{Preprint, September 2026. Pre-registration: OSF \url{https://osf.io/rts6n}.}}

\markboth{Preprint, September 2026}{Abbasi: Inheritance in Distilled Encrypted-Traffic Classifiers}

\maketitle

\begin{abstract}
Knowledge distillation is the standard way to compress encrypted-traffic classifiers for the edge, and almost all such work judges students by accuracy alone. We ask what else a student inherits: unknown-traffic detection, calibration, shortcut reliance, and whether any survives a year of drift. Resemblance proves little on its own, since soft targets also regularise. We therefore distil one 101k-parameter student from two teachers of equal accuracy but different construction, a five-member ensemble and a single wider model, so that following one rather than the other is attributable to it. The design was pre-registered before any test result was seen. We tested ten hypotheses on CESNET-TLS-Year22, a year of real TLS traffic, across 18 test windows over 35 weeks. Two are supported: a student's per-flow unknown-scores shift toward its own teacher, but only at a conventional temperature, not the accuracy-optimal one; and a shortcut-reliant teacher passes its over-confidence to a student that never sees the feature. The drift prediction is reversed under both scores, the gap narrowing rather than widening and the student overtaking under the energy score in two of three replicates, as is the prediction that such a teacher harms its student's detection, which improves slightly. Shortcut reliance is set by model size, not distillation. Under the logit-based scores nothing else transfers: distillation beats neither a temperature-scaled direct student nor label smoothing. Exploratory analysis shows this turns on the scoring rule: with a feature-space detector the teacher detects unknown traffic $0.073$ AUROC better than the direct student, where the energy score sees $0.000$, and the conventional-temperature student inherits most of it. Label smoothing, with no teacher, recovers more. Distillation transfers the teacher's habits; what looks like an inherited ability is available without one.
\end{abstract}

\begin{IEEEkeywords}
Encrypted traffic classification, knowledge distillation, open-set recognition, calibration, concept drift, shortcut learning, pre-registration.
\end{IEEEkeywords}

\IEEEpeerreviewmaketitle

\section{Introduction}

Encrypted-traffic classifiers are getting larger and their deployment targets are not. The models that set the state of the art on public benchmarks are multimodal CNNs, transformers and pre-trained ``traffic foundation models'' with millions to hundreds of millions of parameters~\cite{luxemburk2024year22,etbert,yatc,netfound}, while the places they must run, such as line-rate probes, edge gateways and customer-premises equipment, offer a fraction of a CPU core per flow. Knowledge distillation~\cite{hinton2015} is the standard answer to this mismatch, and a growing body of work applies it to traffic classification~\cite{netclus,merlot,resaware,ciphersight,tnsm2026fedkd}. Almost without exception, that work reports only one thing about the student, its accuracy on a held-out split of the same dataset.

Accuracy is not what makes a traffic classifier deployable. Traffic changes, and the year-long CESNET-TLS-Year22 capture shows classifiers losing tens of points of macro-F1 within months of training~\cite{luxemburk2024year22,ntccrisis}. It also contains things the classifier has never seen, and a model that confidently assigns a new service to an old label is worse than one that abstains~\cite{luxemburk2022reject}. And traffic datasets are full of shortcuts (features that predict the label in the capture but not in the world), which a model may learn instead of the intended signal~\cite{biasseeker}. A large model earns its keep by handling these problems, not by its accuracy alone. Whether a small model distilled from it inherits that handling is the question this paper asks.

The question has a sharper form than ``does the student inherit?''. Soft targets are a regulariser~\cite{yuan2020revisiting,functionalkd}, so much of what distillation appears to transfer could be produced by smoothing the labels, with no teacher involved. To claim that a student inherited something \emph{from its teacher}, one needs a control in which the teacher changes and nothing else does.

We build that control with two teachers that reach the same accuracy by different means: a five-member ensemble and a single wider model of the same family. Each is distilled into the same student. We then test whether each student's per-flow unknown-scores follow its own teacher more closely than they follow the other teacher. The comparison is made against a directly trained student, because the two teachers are not equally ``central''. Any student, distilled or not, correlates more with the ensemble. Subtracting the undistilled student's preference isolates the part attributable to distillation. We apply the same test to the properties operators care about, and we repeat it over a year, because a property inherited at deployment time may not still be inherited three months later.

The teachers studied here sit at the small end of the range just described. They have 2.3M parameters each, 11.3M as an ensemble, and are distilled into a 101k-parameter student. That is the 112-fold compression an operator faces when moving a dataset-scale classifier onto a probe, and it is the regime in which the question can be answered at all, because a teacher-swap control requires training several teachers to equal accuracy and repeating the whole grid at three points in the year. Whether the same conclusions hold when the teacher is a hundred-million-parameter foundation model is a separate question that this design cannot settle.

We pre-registered the design, froze it after a pilot and a validation-only design phase, registered it publicly, and only then computed results on 18 test windows spanning 35 weeks (OSF \url{https://osf.io/rts6n}). Ten hypotheses were tested with Holm correction; two were supported, three came out reversed, and five were not supported.

\subsubsection*{Contributions}
\begin{enumerate}
\item \textbf{A teacher-swap protocol for measuring inheritance.} A distilled student's per-flow unknown-score pattern shifts measurably toward the specific teacher it was distilled from, relative to a directly trained student, and this shift exists only at a conventional distillation temperature. The accuracy-optimal temperature transmits nothing teacher-specific.
\item \textbf{A negative result on what inheritance buys.} Against the cheapest control (the same student trained directly and temperature-scaled), distillation improves no unknown-detection outcome we measured; against label smoothing it buys calibration at the cost of detection. What the $T=4$ student does take from its teacher includes the teacher's poorer detection of near unknowns.
\item \textbf{Drift in the wrong direction for the usual worry.} Over 35 weeks, the 11.3M-parameter teacher ensemble loses unknown-detection ability faster than the 101k student. In two of the three replicates the student becomes the better detector after about three months; in the third the teacher keeps a small lead throughout.
\item \textbf{Shortcuts are a capacity problem, not a distillation problem.} Below full shortcut reliability, small students rely on a planted shortcut about twice as much as the teacher regardless of how they are trained, and distillation does not increase that reliance. A shortcut-reliant teacher does, however, pass its over-confidence through the soft targets alone, to a student that never encounters the feature.
\item \textbf{A released, pre-registered benchmark} on public data: frozen service split, per-flow scores for every model and window, and analysis code with a cluster bootstrap that respects the day\,$\times$\,service dependence structure of traffic data.
\end{enumerate}

The rest of the paper is organised as follows. Section~\ref{sec:related} places the study among work on distillation in traffic classification and on what distillation transfers. Section~\ref{sec:methods} describes the design: the data, the teachers and students, the training conditions and the ten pre-registered hypotheses. Section~\ref{sec:results} reports the confirmatory results by research question. Section~\ref{sec:exploratory} reports the analyses added once those results were in hand, which test the scoring rule, the composition of the teacher swap, the ageing mechanism and the student's capacity. Section~\ref{sec:discussion} discusses what the results mean for an operator and where they disagree with the literature, and Section~\ref{sec:conclusion} concludes.

\section{Related Work}
\label{sec:related}

\subsubsection*{Distillation in traffic classification}
Distillation appears in the traffic literature as a compression tool~\cite{netclus,prunedtrees,futureinternet2026}, a cross-environment transfer mechanism~\cite{resaware,ciphersight}, a self-supervised pre-training objective~\cite{mmaeflowmix}, and a component of federated training~\cite{tnsm2026fedkd}. ResAware~\cite{resaware} is the closest in spirit. It reports that a website-fingerprinting student inherits some of its teacher's calibration and open-world behaviour, and it does measure over time, across a 150-day drift window. What it does not do is separate the teacher's contribution from the regularisation any soft target provides: it tracks how closely the student follows its teacher rather than whether the student's own detection quality beats a teacher-free control, and it uses one teacher on one task, so a shift toward that teacher cannot be attributed to its identity. NetClus~\cite{netclus} distils a traffic transformer and holds out one malware class, on random splits of older datasets. None of these works evaluates the student's unknown-traffic detection or its calibration against a teacher-free control, or its shortcut reliance at all, and none tests whether what the student learned came from the teacher rather than from soft labels in general.

\subsubsection*{What distillation transfers}
Outside networking, the assumption that students inherit their teachers' properties has been examined directly. Stanton et al.~\cite{stanton2021} showed that students often fail to match their teachers' outputs even when they match their accuracy; Ojha et al.~\cite{ojha2023} and Lukasik et al.~\cite{teacherspet} showed that invariances and biases do transfer; Mason-Williams et al.~\cite{functionalkd} found across 22 configurations that functional transfer is weaker and more asymmetric than assumed and reframed distillation as a data-dependent regulariser, echoing the label-smoothing equivalence of Yuan et al.~\cite{yuan2020revisiting}. Most relevant to our temperature result, \emph{Beyond Dark Knowledge}~\cite{beyonddarkknowledge} reports that calibration propagates from teacher to student independently of accuracy and that the temperature governs an accuracy--calibration trade-off; \cite{cud2026,teachercalibration2025} design or study teachers for that purpose. A recent position paper argues that distillation should be judged on the teacher capabilities it preserves rather than the task score it retains~\cite{kdloses2026}; our study is evidence of the kind it calls for. None of this work uses teacher identity as a control, evaluates open-set detection as the inherited property, or measures anything over time.

\subsubsection*{Whose student is it?}
Whether a student's behaviour is specific to its teacher has been studied as forensics: tracing an LLM student to its teacher through lexical fingerprints~\cite{whotaughtyou}, detecting distillation from a given open-weights model~\cite{kddetection2025}, and watermarking against it~\cite{antidistill,tokenprint}. We borrow the question but change its purpose. Teacher identity is here a control for whether a behavioural property, the ranking of flows by how unknown they look, was inherited, and the directly trained student supplies the baseline that forensic work does not need.

\subsubsection*{Open-set and drift-aware traffic classification}
Rejecting unknown services on CESNET data was studied by the dataset authors~\cite{luxemburk2022reject}, and open-world evaluation with leave-one-service-out folds appears in recent darknet work~\cite{darknetopenworld}, without distillation. Drift is documented at week scale in the Year22 dataset itself~\cite{luxemburk2024year22} and is the subject of a growing critique of evaluation practice: redundancy and leakage in benchmarks~\cite{ntccrisis}, label provenance~\cite{soklabels}, and distribution-shift robustness~\cite{unialign,colossus}. Our time-based splits, held-out unknown services, duplicate accounting and pre-registration follow the recommendations of that critique.

\subsubsection*{Shortcuts and compression}
Shortcut learning in encrypted-traffic classification was recently mapped across 19 datasets~\cite{biasseeker}, and spurious-correlation mitigation has entered the traffic literature~\cite{tddm}. That shortcuts pass through distillation is assumed by a wave of mitigation methods in vision and language~\cite{asd2026,saopd2026,iga2026}, and synthetic injected bias features with flip tests have been used to study it in medical imaging~\cite{medshortcutkd}. That compressed models bear a disproportionate share of a model family's brittleness is established for pruning and quantisation~\cite{hooker2020,mtforget,robustnessdistill}. We measure rather than mitigate, on traffic, and we include the condition those works omit, a student that never sees the shortcut and is influenced only through soft targets.

\subsubsection*{Uncertainty-preserving distillation and label smoothing}
Ensemble distribution distillation~\cite{malinin2020endd} and its descendants~\cite{logitendd,selfdistdist,credalendd} report students that match or exceed teacher ensembles at uncertainty estimation; the proxy-Dirichlet formulation we use is due to Ryabinin et al.~\cite{ryabinin2021}. Several studies report that label smoothing harms out-of-distribution detection~\cite{softlabelsood,lsembedding,adaptivels,lsselective}. Our results for both families run the other way; Section~\ref{sec:discussion} addresses the difference.

\section{Experimental Design}
\label{sec:methods}

\begin{figure*}[t]
\centering
\includegraphics[width=\textwidth]{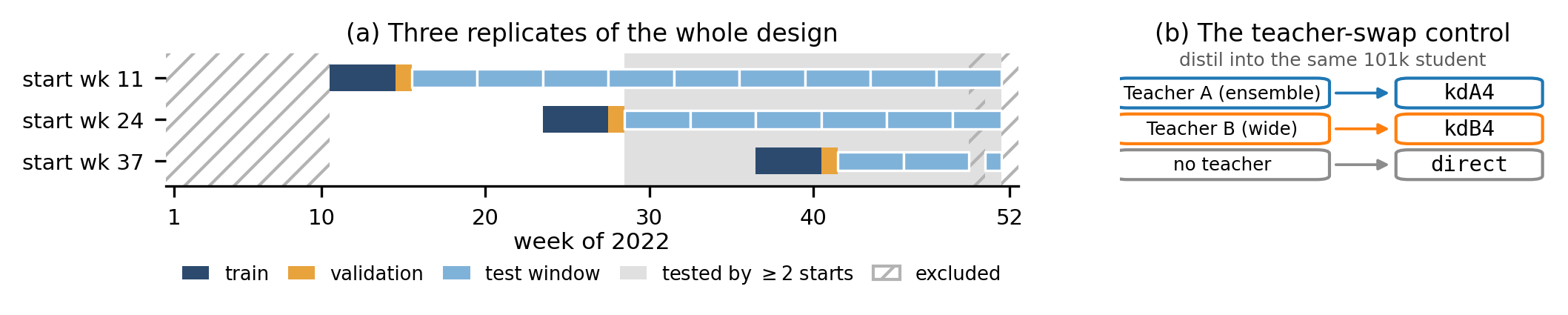}
\caption{The design. (a) The three replicates: four training weeks, one validation week, then consecutive four-week test windows. Shading marks the weeks that more than one start date tests (Section~\ref{sec:overlap}). (b) The teacher-swap control.}
\label{fig:design}
\end{figure*}

The study was pre-registered (OSF \url{https://osf.io/rts6n}, frozen at repository commit \texttt{fa8f718}) after a pilot and a validation-only design phase, and before any test-window result was computed. The 35 weeks over which drift is measured are weeks of the 2022 capture, not of elapsed study time. The confirmatory results were computed after the freeze commit and before the OSF registration was posted, so the freeze precedes them in the public commit history rather than in an assurance, and nothing in the design changed between the two. This section describes the design as registered, and Fig.~\ref{fig:design} summarises it. The same grid of models is trained three times at three points in the year. Within each replicate, two teachers matched on accuracy but built differently are distilled into the same student, and inheritance is measured as the student's own-minus-other correlation with the two teachers, minus the same quantity for a student trained without any teacher. Subtracting that baseline removes the part of the preference that has nothing to do with distillation.

\subsection{Data}

\subsubsection*{Dataset}
CESNET-TLS-Year22~\cite{luxemburk2024year22} contains one year (2022) of TLS flows from the Czech national research network, labelled with 180 web services from the server name. We use the S subset (25M flows) through \texttt{cesnet-datazoo}~0.2.0. Weeks 1--10 are excluded because the flow exporter was updated in week 10, which changed the packet-sequence distribution; weeks 50 and 52 are excluded because they contain less than half the median weekly volume. No service first appears mid-year (177 of 180 are present in every week), so unknown services are simulated by holding them out rather than observed emerging.

\subsubsection*{Inputs}
Each flow is represented by its first 30 packets as three sequences (inter-packet time, direction, size), plus 12 flow statistics and a 32-bin packet-size histogram (44 tabular features). Server name, IP addresses, ports, autonomous system and JA3 fingerprint are not used, so the classifier sees what it would see under Encrypted Client Hello. Sequences and statistics are scaled with transforms fitted on each training window only.

\subsubsection*{Known and unknown services}
A fixed split, generated once with seed 2022 before any model was trained, designates 102 services as known. Unknown services are held out and differ between validation and test, so no threshold or design choice made on validation unknowns can be tuned to the test unknowns. Each unknown set is divided into \emph{far} unknowns, whole categories held out (validation: Music and Videoconferencing, 8 services; test: Search and Social, 11 services), and \emph{near} unknowns, individual services whose category remains known (validation: 5; test: 8). The remaining 46 services are excluded, mostly for insufficient or inconsistent volume.

\subsubsection*{Time structure}
Three training start dates give three replicates of the whole design: training on weeks 11--14, 24--27 and 37--40 (1.89M, 1.37M and 1.35M known flows), validation on the following week (15, 28, 41; 200k known flows plus every flow of the validation unknown services), and testing on consecutive 4-week windows from the week after validation to week 52, skipping the excluded weeks. This yields 9, 6 and 3 test windows, 18 in total, each a stratified sample of 100k known and 100k test-unknown flows, at 3.5 to 35.3 weeks after the end of training. The one exception is start date 37's last window, which the end of the year truncates to a single week and which supplies 43,579 unknown flows. Each test flow carries its day, its packet count, and a flag for whether its exact 30-packet sequence occurs in its training window (a 64-bit hash of the scaled sequence; 1.8\% of test flows).

\subsection{Models}

\textbf{Teacher A} is an ensemble of five \texttt{mm\_cesnet\_v2} networks~\cite{luxemburk2024year22}, the dataset authors' multimodal CNN (2.26M parameters each, seeds 0--4). Its prediction is the mean of the members' softmax outputs; its energy score is the mean of the members' energies. Averaging the members' energies, rather than taking the energy of the mean logits, keeps the ensemble's score on the same scale as a single member's, so that a teacher--student comparison is not confounded by an aggregation that shifts the score distribution. The choice was fixed in the pre-registration. Teachers A and B differ in ensembling and width but belong to the same architecture family, a limitation of the swap that Section~\ref{sec:exploratory} addresses with a transformer teacher and with two single-member teachers.

\textbf{Teacher B}, the teacher-swap control, is a single wider network of the same family (three convolutional blocks of 400/600/600 channels, two 450-unit flow-statistics layers, a 1,200-unit shared layer; 10.3M parameters). It must be within 2 macro-F1 points of Teacher A on the validation week of each start date, and it was within 0.2 points at all three.

The \textbf{student} is a multimodal 1D-CNN of 101,190 parameters: three convolutions (48, 96, 96 channels; kernels 5, 5, 3) over the packet sequences, pooled by mean and max; a 64-unit layer over the flow statistics; a 128-unit shared layer with dropout 0.1; and a linear classifier. It is 22$\times$ smaller than one teacher member and 112$\times$ smaller than the ensemble.

\textbf{Baselines} (validation week only): XGBoost on the 44 flow statistics (500k training flows, 300 trees, depth 8) with the top-two margin as unknown-score, and $k$-nearest-neighbours on the packet sequences (300k reference flows, $k=10$) with the negative $k$-th distance.

\subsection{Training Conditions}

Every network is trained with AdamW (learning rate $10^{-3}$, weight decay $10^{-4}$), a one-cycle schedule with 10\% warm-up, batch size 1,024 and bf16 autocast; teachers for 10 epochs and students for 20. The kept checkpoint is the epoch with the lowest cross-entropy on the validation week's known flows. Table~\ref{tab:conditions} lists the student conditions, each trained with three seeds. All students share the 101k-parameter architecture and this recipe, and differ only as the table states. Hinton distillation minimises $\alpha T^2\,\mathrm{KL}(p^{\text{teacher}}_T \,\|\, p^{\text{student}}_T) + (1-\alpha)\,\mathrm{CE}$ with $\alpha=0.9$, so the conditions differ only in the teacher and the temperature. Ensemble distribution distillation uses proxy-Dirichlet targets~\cite{ryabinin2021} with the precision capped at $10^4$ and a reverse-KL loss; maximum-likelihood distribution distillation was unstable with 102 classes in the pilot and was replaced before freezing. The lower half of the table lists conditions added after the confirmatory results were known; they are exploratory throughout and appear only in Section~\ref{sec:exploratory}.

\begin{table}[t]
\centering
\caption{Student training conditions. The upper block is pre-registered and runs at all three start dates; the lower block is exploratory, added after the confirmatory results were known.}
\label{tab:conditions}
\footnotesize
\setlength{\tabcolsep}{4pt}
\begin{tabular}{@{}llp{38mm}@{}}
\toprule
Condition & Teacher & Objective \\
\midrule
\kd{direct}   & ---              & cross-entropy \\
\kd{directTS} & ---              & \kd{direct}, then post-hoc temperature scaling \\
\kd{ls}       & ---              & label smoothing, $\varepsilon = 0.05$ \\
\kd{kdA}      & A                & Hinton KD, $T=1$ (accuracy-tuned) \\
\kd{kdB}      & B                & Hinton KD, $T=1$ \\
\kd{kdA4}     & A                & Hinton KD, $T=4$ (conventional) \\
\kd{kdB4}     & B                & Hinton KD, $T=4$ \\
\kd{enddA}    & A's five members & proxy-Dirichlet EnDD \\
\midrule
\kd{kdM0}     & A, member 0 only & Hinton KD, $T=4$ \\
\kd{kdM1}     & A, member 1 only & Hinton KD, $T=4$ \\
\kd{kdC}      & C (transformer)  & Hinton KD, $T=4$; start date 11 \\
\kd{hardA}    & A                & cross-entropy on the teacher's top-1 labels \\
\kd{kdF}      & A, member 0 only & similarity-preserving feature KD; start date 11 \\
\kd{\_w16}, \kd{\_w96} & as above & suffixes marking \kd{direct} and \kd{kdA4} retrained at student widths 16 and 96; start date 11 \\
\bottomrule
\end{tabular}
\end{table}

\subsubsection*{Tuning}
One tuning pass on start date 11 (validation week, one seed) chose the KD temperature and weight from $\{1,2,4\}\times\{0.5,0.9\}$ and the smoothing strength from $\{0.05,0.1,0.2\}$, by validation macro-F1 alone; no unknown-detection or calibration metric was computed during tuning. The accuracy-optimal temperature was $T=1$. Because a 96\%-accurate teacher's targets at $T=1$ are nearly one-hot, the validation grid showed \kd{kdA} and \kd{kdB} at $T=1$ to be indistinguishable from \kd{direct} in their per-flow scores, leaving nothing for a teacher-swap test to detect. The $T=4$ arm was therefore added, before freezing, as the arm on which inheritance (H1) and the shortcut experiment are tested, while the tuned arm carries the matched-accuracy comparisons (H2, H5); both arms are reported throughout. The same pass showed that 10 epochs left students unconverged (macro-F1 rose by 1.5 points at 20), which fixed the student budget at 20 epochs. Post-hoc temperatures for every model are fitted by minimising NLL on the validation week's known flows.

\subsection{Outcomes}

For each model, window and flow group we report closed-set macro-F1 over the known services; unknown-detection AUROC (area under the receiver operating characteristic curve) and FPR at 95\% TPR with two unknown-scores, the energy score (log-sum-exp of the logits) and the maximum softmax probability (MSP), treating known flows as positives; the open-set classification rate (OSCR); calibration as expected calibration error (ECE, 15 bins), negative log-likelihood (NLL) and Brier score, before and after temperature scaling; and the area under the risk--coverage curve (AURC). Per-flow inheritance is measured by the Spearman correlation of a student's energy scores with each teacher's. Everything is computed for all flows and for flows with at least five packets, against all, near and far unknown services.

\subsection{Hypotheses and Analysis}
\label{sec:analysis}

The unit of analysis is one start date $\times$ one test window (18 units). A per-unit statistic is the mean over the three student seeds; a pooled estimate is the unweighted mean over units. Uncertainty comes from 1,000 bootstrap resamples that redraw day\,$\times$\,service clusters with replacement, \emph{globally}, so that a cluster appearing in the windows of several start dates receives the same multiplicity everywhere, and independently redraw the three seeds per start date; resampled clusters enter every metric as integer flow weights. Intervals are 95\% percentile intervals; $p$-values are one-sided, $(1+\#\{\text{resampled estimate}\le 0\})/1001$, and Holm-corrected across ten hypotheses. A hypothesis with several components is supported only if every component is, so its $p$-value is its largest component $p$-value (intersection--union).

\begin{itemize}
\item \textbf{H1 (teacher-specific inheritance).} For the $T=4$ students, the own-minus-other difference in Spearman correlation with the two teachers, minus the same difference for the \kd{direct} student; both \kd{kdA4} (toward A) and \kd{kdB4} (toward B) must exceed zero. The baseline is necessary because every student correlates more with the ensemble teacher regardless of training.
\item \textbf{H2 (beyond regularisation)}, once per score: \kd{kdA} exceeds each of \kd{ls} and \kd{directTS} in unknown-detection AUROC and has lower post-hoc NLL than each; interpreted as a matched-accuracy comparison only while the macro-F1 difference is at most 1 point (it was 0.99 and 0.08).
\item \textbf{H3 (decay)}, once per score: positive slope of the per-unit Teacher~A\,$-$\,\kd{kdA} AUROC gap on weeks since training, in an OLS fit with one intercept per start date.
\item \textbf{H4a, H4b1, H4b2 (shortcuts).} A one-hot feature with 8 values, equal to the class index modulo 8 with probability $\rho$ and random otherwise, is appended to the flow statistics during training ($\rho\in\{0,0.5,0.9,1.0\}$, three seeds, start date 11, its validation week). Reliance is macro-F1 with the feature aligned to the true class minus macro-F1 with it shifted to a wrong value. H4a: with the feature visible to both, the KD student ($T=4$) relies on it more than the \kd{direct} student at $\rho\in\{0.9,1.0\}$. H4b1 and H4b2: a student that never sees the feature, distilled from a teacher trained at $\rho\in\{0.9,1.0\}$, has higher ECE (H4b1) and lower energy AUROC (H4b2) than one distilled from the $\rho=0$ teacher of the same seed. Paired one-sided $t$-tests over the six ($\rho$, seed) pairs.
\item \textbf{H5 (distribution distillation)}, once per score: \kd{enddA} exceeds \kd{kdA} in unknown-detection AUROC.
\end{itemize}

Two sensitivity analyses repeat the seven test-window hypotheses with exact duplicates of training flows removed and with flows of fewer than five packets removed; they are reported but are outside the Holm family. They do not cover H4a, H4b1 and H4b2, which are computed on the shortcut experiment's validation week, where neither filter is defined. The pilot gate (Teacher A must beat the \kd{direct} student by $\ge 2$ macro-F1 points or $\ge 0.02$ energy AUROC on the validation week) was passed with a 6.2-point macro-F1 gap; the energy-AUROC gap was 0.018, just under its own threshold, so the gate turned on accuracy alone.

\section{Confirmatory Results}
\label{sec:results}

All numbers come from the pre-registered test windows unless marked otherwise. Unknown-scores are oriented so that higher means ``more likely known''; ``gap'' means teacher minus student.

\subsection{Setting and Headline Numbers}

Table~\ref{tab:headline} averages over all 18 windows, and much of what follows turns on its columns. The teachers are 4.8 to 7.1 macro-F1 points more accurate than any student, but their advantage in unknown detection is small, 0.001 energy AUROC over \kd{kdA} (CI $-0.003$ to $0.005$) and 0.025 with MSP (CI 0.023 to 0.026). The accuracy-tuned distillation arm ($T=1$) is indistinguishable from the directly trained student on every column. The two teachers are well matched, Teacher B trailing Teacher A by 0.55 macro-F1 points on average across windows (maximum 1.6), inside the pre-registered tolerance, so the teacher-swap comparison is between equals.

\begin{table*}[t]
\centering
\caption{Test-window averages (18 windows, all flows, all test-unknown services). Higher is better except FPR, NLL and AURC. NLL is post-hoc (temperature-scaled).}
\label{tab:headline}
\begin{tabular}{llrrrrrrr}
\toprule
Model & Params & Macro-F1 & AUROC$_{\text{energy}}$ & AUROC$_{\text{MSP}}$ & FPR@95 & OSCR & NLL & AURC \\
\midrule
Teacher A (5$\times$ mm\_cesnet\_v2) & 11.3M & 0.860 & 0.836 & 0.868 & 0.626 & 0.789 & 0.447 & 0.012 \\
Teacher B (wide)                     & 10.3M & 0.855 & 0.840 & 0.851 & 0.614 & 0.791 & 0.480 & 0.014 \\
\midrule
\kd{direct}           & 101k & 0.806 & 0.835 & 0.844 & 0.674 & 0.768 & 0.609 & 0.020 \\
\kd{directTS}         & 101k & 0.806 & 0.835 & 0.845 & 0.676 & 0.767 & ---   & 0.020 \\
\kd{ls} ($\varepsilon=0.05$) & 101k & 0.797 & 0.814 & 0.836 & 0.683 & 0.747 & 0.561 & 0.023 \\
\kd{kdA} ($T=1$)      & 101k & 0.806 & 0.834 & 0.843 & 0.667 & 0.767 & 0.601 & 0.020 \\
\kd{kdB} ($T=1$)      & 101k & 0.806 & 0.831 & 0.840 & 0.674 & 0.764 & 0.602 & 0.020 \\
\kd{kdA4} ($T=4$)     & 101k & 0.790 & 0.809 & 0.824 & 0.678 & 0.739 & 0.565 & 0.023 \\
\kd{kdB4} ($T=4$)     & 101k & 0.789 & 0.810 & 0.822 & 0.669 & 0.741 & 0.582 & 0.023 \\
\kd{enddA}            & 101k & 0.801 & 0.742 & 0.835 & 0.780 & 0.679 & 0.563 & 0.022 \\
\bottomrule
\end{tabular}
\end{table*}

\begin{table*}[t]
\centering
\caption{The ten pre-registered hypotheses. Estimates are pooled over the 18 test windows; intervals are 95\% percentile intervals from 1,000 day$\times$service cluster bootstrap resamples; $p$-values are Holm-corrected over these ten hypotheses and no others.}
\label{tab:hypotheses}
\footnotesize
\begin{tabular}{@{}l l p{70mm} r l@{}}
\toprule
\# & Hypothesis & Estimate (95\% CI) & Holm $p$ & Outcome \\
\midrule
H1 & inheritance is teacher-specific & \kd{kdA4} $+0.007$ (0.005, 0.009); \kd{kdB4} $+0.026$ (0.024, 0.028) & 0.009 & \textbf{supported} \\
H2[energy] & KD beats \kd{ls}, \kd{directTS} at matched accuracy & AUROC vs \kd{ls} $+0.021$, vs \kd{directTS} $0.000$; NLL vs \kd{ls} $-0.040$, vs \kd{directTS} $+0.008$ & 1.0 & not supported \\
H2[MSP] & as H2, MSP & AUROC vs \kd{ls} $+0.006$, vs \kd{directTS} $-0.002$ & 1.0 & not supported \\
H3[energy] & gap grows with time & $-0.0004$ per week ($-0.0007$, $-0.0001$) & 1.0 & reversed \\
H3[MSP] & as H3, MSP & $-0.0003$ per week ($-0.0004$, $-0.0001$) & 1.0 & reversed \\
H4a & KD student relies on a visible shortcut more & $-0.032$ (lower bound $-0.071$) & 1.0 & not supported \\
H4b1 & shortcut-reliant teacher transfers over-confidence & ECE $+0.0086$ (lower bound 0.0061) & 0.005 & \textbf{supported} \\
H4b2 & \ldots and degrades unknown detection & $-0.020$ (detection improves) & 1.0 & reversed \\
H5[energy] & EnDD keeps more than Hinton KD & $-0.092$ ($-0.099$, $-0.084$) & 1.0 & not supported \\
H5[MSP] & as H5, MSP & $-0.008$ ($-0.009$, $-0.006$) & 1.0 & not supported \\
\bottomrule
\end{tabular}
\end{table*}

Table~\ref{tab:hypotheses} gives the ten pre-registered tests. All of them are one-sided, so H4's rows report the one-sided lower confidence bound from paired $t$-tests over six ($\rho$, seed) pairs, and the two-sided percentile intervals of the other rows may be read the same way. With 1,000 resamples the smallest attainable one-sided $p$ is $1/1001 = 0.000999$, so a bootstrap hypothesis at that floor reaches $0.010$ when it ranks first in a family of ten and $0.009$ when it ranks second, which is where H1's value comes from; rows shown as 1.0 are bounded rather than precisely measured. The $T=4$ arm that H1 tests (\kd{kdA4}, \kd{kdB4}) was added before the freeze but after the validation weeks had shown that distillation at the accuracy-optimal temperature transfers nothing teacher-specific; the $T=1$ arm is reported beside it in Fig.~\ref{fig:inheritance}.

The two sensitivity analyses (exact duplicates of training flows removed, 1.8\% of test flows; flows with at least five packets only, 94.0\% of test flows) reach the same decision on each of the seven hypotheses they cover, with every $p$-value within 0.1 of the primary analysis. The three shortcut hypotheses are not among them, for the reason given in Section~\ref{sec:analysis}.

\subsection{RQ1: What the Student Inherits}

\subsubsection*{The score pattern transfers, and it is teacher-specific (H1)}
\begin{figure*}[t]
\centering
\includegraphics[width=\textwidth]{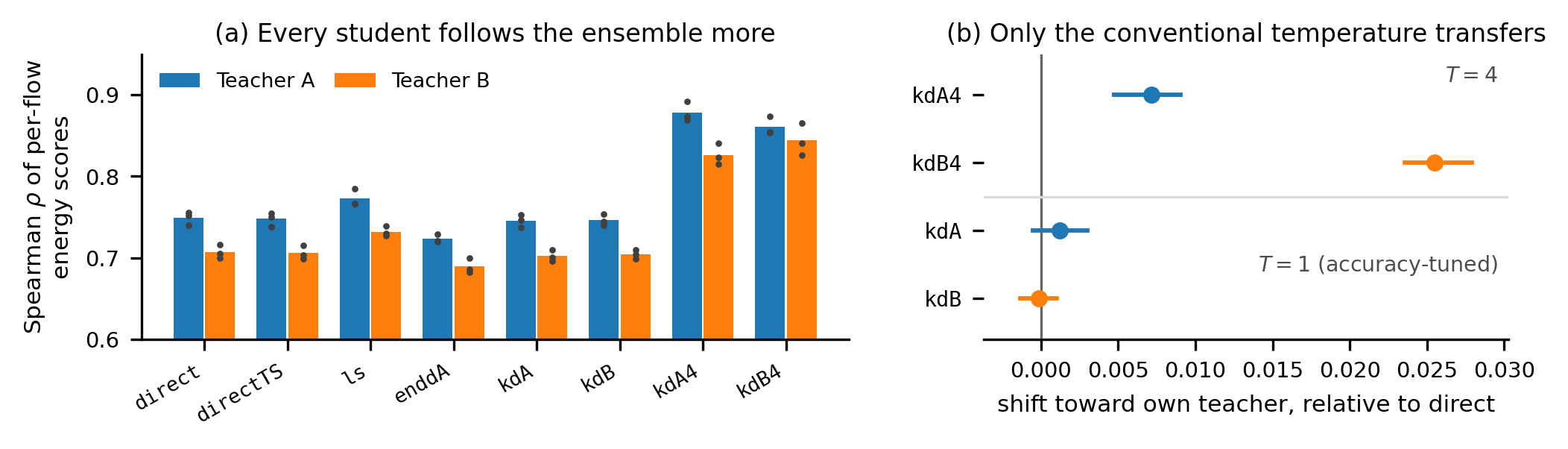}
\caption{Inheritance of the per-flow score pattern. (a) Spearman correlation of a student's energy scores with each teacher's, averaged within each start date (dots) and then across the three. (b) The quantity H1 tests, pooled over the 18 test windows: each distilled student's own-minus-other correlation difference, minus that of \kd{direct}, with 95\% cluster-bootstrap intervals.}
\label{fig:inheritance}
\end{figure*}

Every student correlates more with Teacher A than with Teacher B, including the student that was never distilled (Spearman $\rho$ of per-flow energy scores: 0.749 against 0.707 for \kd{direct}; Fig.~\ref{fig:inheritance}(a)). The reason is that Teacher A is an ensemble, so its scores are the more ``average'' of the two. A raw own-minus-other comparison would therefore credit this artefact to distillation. H1 avoids that by measuring each distilled student's preference \emph{relative to the direct student}. On that measure both conventional-KD students move toward their own teacher, \kd{kdA4} by $+0.007$ and \kd{kdB4} by $+0.026$ (Fig.~\ref{fig:inheritance}(b)). Both effects are small in absolute terms, but they are estimated from 3.5 million test flows and survive Holm correction ($p=0.009$).

The two arms are not equally strong. The \kd{kdB4} shift is positive in all 18 windows, from 0.017 to 0.033. The \kd{kdA4} shift is positive in the nine windows of start dates 24 and 37 (0.011 to 0.023), but indistinguishable from zero in the nine windows of start date 11 ($-0.007$ to $+0.005$). We have no pre-registered explanation for this asymmetry and report it as a limitation of H1's support.

The accuracy-tuned arm inherits nothing. Its shifts are $+0.001$ (CI $-0.001$, 0.003) for \kd{kdA} and $-0.000$ (CI $-0.001$, 0.001) for \kd{kdB}. This is a null result about distillation at $T=1$ from a 96\%-accurate teacher, not about distillation in general. Such a teacher's soft targets are almost one-hot, so they carry little information beyond the label. The temperature therefore controls how much teacher-specific signal reaches the student, and the temperature our tuning procedure selected for accuracy is the one that transmits none of it.

\subsubsection*{Detection quality and calibration do not transfer (H2, H5)}
Table~\ref{tab:headline} already shows the result. \kd{kdA} and \kd{directTS} differ by $0.00003$ in energy AUROC, by $0.002$ in MSP AUROC, and by less than $0.001$ in OSCR, AURC and macro-F1. Temperature-scaling a directly trained student, which costs one scalar fitted on validation data, matches distillation on every unknown-detection outcome we measured. Against label smoothing the picture is mixed and does not favour distillation overall, since \kd{kdA} detects unknowns better ($+0.021$ energy AUROC, $+0.006$ MSP) but is worse calibrated (post-hoc NLL 0.601 vs 0.561). Because H2 requires distillation to win on both outcomes against both controls, it is not supported for either score. The ensemble-distribution-distillation student is the worst unknown detector in the grid with the energy score (0.742, a 0.092 deficit to \kd{kdA}) and slightly worse with MSP ($-0.008$); H5 is not supported for either score.

\subsubsection*{Near and far unknowns}
Table~\ref{tab:nearfar} splits detection by the distance of the unknown services. The small direct student is a \emph{better} detector of near unknowns (services from a known category) than either teacher with the energy score, and the teachers' advantage is confined to far unknowns and to the MSP score. Distillation at $T=4$ moves the student toward the teacher's profile. Relative to \kd{direct}, \kd{kdA4} loses 0.059 energy AUROC on near unknowns and lands on Teacher B's value to three decimals. Inheriting the teacher's score pattern therefore includes inheriting the teacher's weakness.

\begin{table}[t]
\centering
\caption{Unknown detection by distance of the unknown services (test windows; energy AUROC / MSP AUROC). The three conditions sharing a row agree to within 0.002 on every entry; the row gives \kd{direct}.}
\label{tab:nearfar}
\begin{tabular}{lcc}
\toprule
Model & Near unknowns & Far unknowns \\
\midrule
Teacher A & 0.805 / 0.836 & 0.840 / 0.872 \\
Teacher B & 0.788 / 0.794 & 0.848 / 0.859 \\
\kd{direct}, \kd{directTS}, \kd{kdA} & 0.846 / 0.824 & 0.833 / 0.847 \\
\kd{ls}   & 0.795 / 0.809 & 0.816 / 0.840 \\
\kd{kdA4} & 0.788 / 0.794 & 0.812 / 0.829 \\
\kd{enddA} & 0.695 / 0.816 & 0.749 / 0.838 \\
\bottomrule
\end{tabular}
\end{table}

\subsection{RQ2: The Gap over Time (H3)}

\begin{figure*}[t]
\centering
\includegraphics[width=\textwidth]{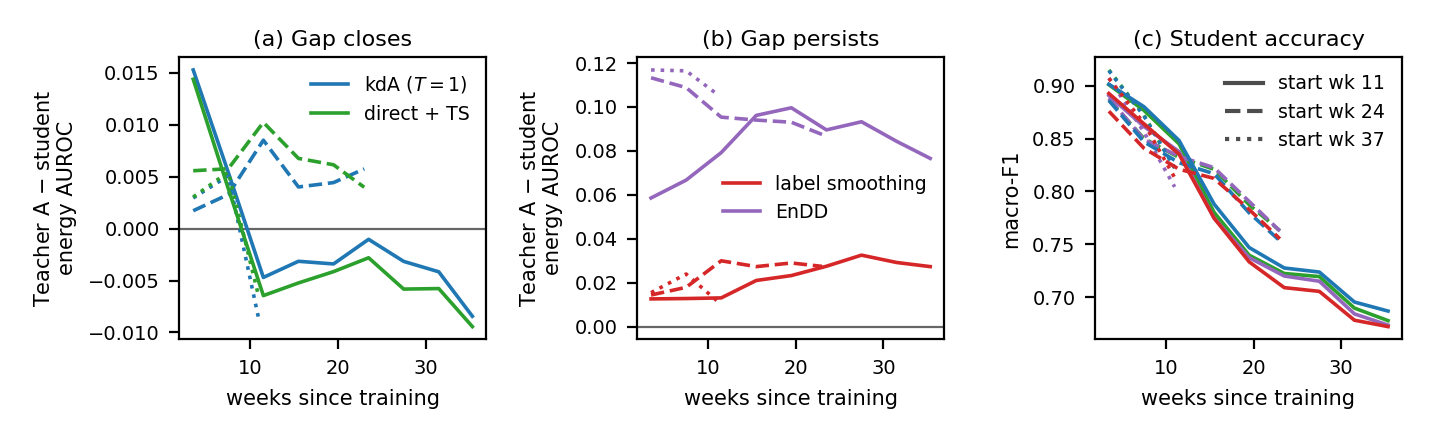}
\caption{Teacher-minus-student energy AUROC against weeks since the end of training, per condition and start date: (a) the conditions whose gap closes and crosses zero; (b) those whose gap does not. The panels use different vertical scales. (c) Student macro-F1 over the same axis. Line style marks the start date.}
\label{fig:gap}
\end{figure*}

We pre-registered that the teacher--student gap would grow with weeks since training. It shrinks (Fig.~\ref{fig:gap}(a)). The fixed-effects slope is $-0.0004$ energy AUROC per week (CI $-0.0007$ to $-0.0001$) and $-0.0003$ per week with MSP, both opposite in sign to H3. Table~\ref{tab:time} follows start date 11. Both models degrade over the year, and steeply, with closed-set macro-F1 for every student falling from about 0.90 to about 0.69 over 35 weeks (Fig.~\ref{fig:gap}(c)). But the 11.3M-parameter ensemble loses energy AUROC faster than the 101k student, and from roughly the third month onwards the student is the better unknown detector at this start date. The crossing is not universal: it also occurs at start date 37, but at start date 24 the teacher leads in all six of its windows, by $0.002$ to $0.009$, so pooled over all 18 windows the teacher is still nominally ahead ($+0.001$, CI $-0.003$ to $0.005$). What the pooled data establish is the reversed slope, not the crossing. The gap narrows under MSP as well, $-0.0003$ per week, so the two scores agree on the direction; what they disagree on is whether the gap reaches zero. Under MSP the teacher holds a lead of about 0.026 for the first thirty weeks and 0.011 in the last window, never losing it, whereas under the energy score the student overtakes. With neither score does the gap widen. The label-smoothing student is the exception in the other direction, its gap to the teacher growing from 0.013 to 0.027 over start date 11.

\begin{table}[t]
\centering
\caption{Start date 11: energy AUROC against all test-unknown services, and the student's closed-set macro-F1, by weeks since the end of training. Six of this start date's nine windows are shown.}
\label{tab:time}
\begin{tabular}{rrrrr}
\toprule
Weeks & Teacher A & \kd{kdA} & Gap & \kd{kdA} macro-F1 \\
\midrule
3.5  & 0.869 & 0.854 & $+0.015$ & 0.902 \\
7.5  & 0.863 & 0.858 & $+0.005$ & 0.880 \\
11.5 & 0.841 & 0.846 & $-0.005$ & 0.848 \\
19.5 & 0.815 & 0.818 & $-0.003$ & 0.747 \\
27.5 & 0.803 & 0.806 & $-0.003$ & 0.724 \\
35.3 & 0.784 & 0.793 & $-0.008$ & 0.687 \\
\bottomrule
\end{tabular}
\end{table}

\subsection{RQ3: Shortcuts (H4a, H4b1, H4b2)}

\begin{figure}[t]
\centering
\includegraphics[width=\columnwidth]{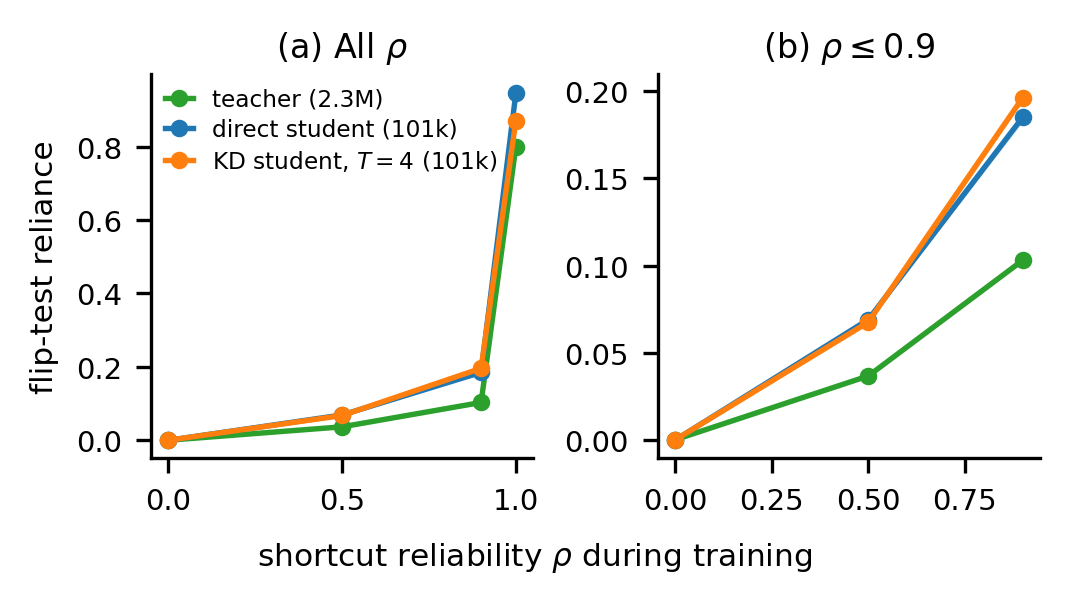}
\caption{Flip-test reliance against the planted feature's reliability $\rho$ during training, for the teacher and the two students that see it: (a) the full range; (b) $\rho \leq 0.9$ on its own scale. Lines are the mean of three seeds, bands the seed range.}
\label{fig:reliance}
\end{figure}

The shortcut experiment plants a one-hot feature that agrees with the class with probability $\rho$ during training and measures reliance as the macro-F1 lost when the feature is flipped to a wrong value (Table~\ref{tab:reliance}, Fig.~\ref{fig:reliance}). Distillation does not increase shortcut reliance. At $\rho\in\{0.9,1.0\}$ the KD student relies on the feature 0.032 \emph{less} than the direct student (H4a not supported), a difference driven by $\rho=1$, where the direct student collapses almost completely. What the table does show is a capacity effect. At every $\rho$ below 1, both 101k students rely on the shortcut about twice as much as the 2.3M-parameter teacher (0.185--0.196 vs 0.103 at $\rho=0.9$), whichever way they were trained.

In the teacher-only setting the student never sees the feature and can be influenced only through the soft targets. A teacher trained with a reliable shortcut ($\rho\in\{0.9,1.0\}$) produces a student with higher ECE than a shortcut-free teacher does ($+0.0086$, Holm $p=0.005$; H4b1 supported). Over-confidence transfers even when the feature that caused it does not. The same student's energy AUROC is \emph{higher}, not lower ($+0.020$; H4b2 reversed), and its macro-F1 is unchanged. Whatever the shortcut-reliant teacher's soft targets carry, it sharpens the student's confidence without damaging its ranking of unknown flows.

\begin{table}[t]
\centering
\caption{Flip-test reliance on the planted shortcut (macro-F1 aligned $-$ flipped; validation week of start date 11, mean of 3 seeds).}
\label{tab:reliance}
\begin{tabular}{cccc}
\toprule
$\rho$ & Teacher (2.3M) & \kd{direct} student & KD student ($T=4$) \\
\midrule
0   & 0.000 & 0.000 & 0.000 \\
0.5 & 0.037 & 0.069 & 0.068 \\
0.9 & 0.103 & 0.185 & 0.196 \\
1.0 & 0.800 & 0.945 & 0.870 \\
\bottomrule
\end{tabular}
\end{table}

\subsection{Baselines and Deployment Cost}

This subsection is exploratory, and is placed here because it sets the scene for the confirmatory results rather than because it shares their status. The pre-registration lists the baselines and the deployment measurements among its planned exploratory analyses (\S9), and the re-tuned XGBoost configuration was added later still, at peer review (\S11). None of it enters the Holm family.

On the validation week, XGBoost on the 44 flow statistics reaches macro-F1 0.883 and energy-margin AUROC 0.855, against 0.964 and 0.837 for Teacher A: the gradient-boosted model is 8 points less accurate but a slightly better detector of unknown services, and clearly better on near unknowns (0.875 vs 0.828). $k$-NN on packet sequences reaches 0.712 / 0.745. Table~\ref{tab:cpu} reports inference cost on a laptop CPU (AMD Ryzen 7 5800H) as a stand-in for an edge device, with weights measured as the serialised model size; the teachers could not be exported to ONNX Runtime. The student in ONNX Runtime processes a flow in 0.08\,ms on one thread, 230$\times$ faster than the teacher ensemble and 43$\times$ faster than one teacher member, in 0.4\,MB of weights. XGBoost is not the cheap alternative it is often taken to be at this class count: 300 boosting rounds over 102 classes is 30,600 trees, a 244\,MB model and 42\,ms per flow, 500$\times$ the student's latency; it recovers only in large batches on all cores. That configuration was chosen for accuracy, not deployment, so we also re-tuned it to a deployable size (150 rounds, depth 6). The smaller model costs 2.9 macro-F1 points (0.854 against 0.883) and 0.029 energy-margin AUROC (0.826 against 0.855), and is 3.3$\times$ smaller at 73\,MB. It was timed in a separate session, at 14.4\,ms per flow against 0.072\,ms for the student re-timed alongside it, so the two configurations should be compared through that session's own ratio: 200$\times$ the student's latency and 180$\times$ its weight footprint. The comparison's direction is unchanged while its magnitude is fairer.

\begin{table*}[t]
\centering
\caption{Inference cost on a laptop CPU (AMD Ryzen 7 5800H, 8 cores) as a stand-in for an edge device. Latency is per flow at batch size 1 on one thread; throughput is at batch size 1,024. All rows come from one timing session, so they are comparable with each other; the re-tuned XGBoost of Section~\ref{sec:results} was timed separately and is quoted against its own session's student.}
\label{tab:cpu}
\begin{tabular}{lrrrrrr}
\toprule
Model & Parameters & Weights (MB) & Latency p50 (ms) & Latency p99 (ms) & Flows/s, 1 thread & Flows/s, all cores \\
\midrule
Student, ONNX Runtime & 101k & 0.4 & 0.082 & 0.160 & 21,201 & --- \\
Student, PyTorch & 101k & 0.4 & 0.501 & 1.026 & 10,604 & 39,418 \\
Teacher A, one member & 2.3M & 9.1 & 3.560 & 5.348 & 812 & 3,875 \\
Teacher A, 5-member ensemble & 11.3M & 45.4 & 19.093 & 27.357 & 161 & 789 \\
Teacher B & 10.3M & 41.3 & 12.262 & 17.213 & 218 & 1,153 \\
XGBoost (flow statistics) & --- & 244.4 & 41.606 & 53.011 & 1,369 & 11,078 \\
\bottomrule
\end{tabular}
\end{table*}

\section{Exploratory Analyses}
\label{sec:exploratory}

The analyses in this section were added after the confirmatory results were known. None of them is pre-registered, none shares a multiple-comparison family with the ten hypotheses of Table~\ref{tab:hypotheses}, and none alters a number reported above. The frozen analysis code was re-run alongside them and reproduces the confirmatory tables exactly. Intervals come from the same day$\times$service cluster bootstrap, resampling training seeds in the same way, so that they can be read against the confirmatory intervals.


\begin{table*}[t]
\centering
\caption{Detection advantage of the exploratory conditions. A row naming one condition is that condition minus \kd{direct}; a row naming two is their paired difference. A dash marks a score that was not computed. Intervals are 95\% cluster bootstrap intervals over the 18 test windows, except for cells marked $\dagger$, which pool over the nine windows of start date 11 because the condition was trained, or in the feature-space columns scored, there alone. A marked cell and an unmarked one are measured on different calendars and should not be read against each other.}
\label{tab:logitscores}
\footnotesize
\setlength{\tabcolsep}{4pt}
\begin{tabular}{@{}lrrrr@{}}
\toprule
Comparison & Energy & MSP & Mahalanobis & Feature $k$-NN \\
\midrule
\kd{direct\_w16} & $-0.024^{\dagger}$ \tiny($-0.033$, $-0.018$) & $-0.008^{\dagger}$ \tiny($-0.017$, $+0.005$) & --- & --- \\
\kd{direct\_w96} & $+0.002^{\dagger}$ \tiny($-0.002$, $+0.005$) & $-0.003^{\dagger}$ \tiny($-0.006$, $-0.001$) & --- & --- \\
\kd{directTS\_w16} & $-0.024^{\dagger}$ \tiny($-0.033$, $-0.018$) & $-0.007^{\dagger}$ \tiny($-0.017$, $+0.005$) & --- & --- \\
\kd{directTS\_w96} & $+0.001^{\dagger}$ \tiny($-0.003$, $+0.004$) & $+0.000^{\dagger}$ \tiny($-0.002$, $+0.003$) & --- & --- \\
\kd{kdA4\_w16} & $-0.054^{\dagger}$ \tiny($-0.062$, $-0.045$) & $-0.051^{\dagger}$ \tiny($-0.054$, $-0.048$) & --- & --- \\
\kd{kdA4\_w96} & $-0.016^{\dagger}$ \tiny($-0.020$, $-0.012$) & $-0.015^{\dagger}$ \tiny($-0.018$, $-0.013$) & --- & --- \\
\kd{kdM0} & $-0.028$ \tiny($-0.033$, $-0.024$) & $-0.027$ \tiny($-0.030$, $-0.024$) & $+0.078^{\dagger}$ \tiny($+0.061$, $+0.089$) & $+0.027^{\dagger}$ \tiny($+0.016$, $+0.037$) \\
\kd{kdM1} & $-0.027$ \tiny($-0.031$, $-0.023$) & $-0.026$ \tiny($-0.027$, $-0.024$) & $+0.071^{\dagger}$ \tiny($+0.050$, $+0.091$) & $+0.022^{\dagger}$ \tiny($+0.008$, $+0.033$) \\
\kd{kdC} & $+0.004^{\dagger}$ \tiny($+0.001$, $+0.008$) & $-0.031^{\dagger}$ \tiny($-0.033$, $-0.029$) & --- & --- \\
\kd{hardA} & $-0.003$ \tiny($-0.006$, $-0.001$) & $-0.002$ \tiny($-0.004$, $-0.001$) & $+0.018^{\dagger}$ \tiny($+0.004$, $+0.033$) & $+0.017^{\dagger}$ \tiny($+0.005$, $+0.029$) \\
\kd{kdF} & $+0.003^{\dagger}$ \tiny($-0.001$, $+0.007$) & $+0.001^{\dagger}$ \tiny($-0.001$, $+0.002$) & $+0.014^{\dagger}$ \tiny($-0.010$, $+0.033$) & $+0.018^{\dagger}$ \tiny($+0.001$, $+0.028$) \\
\kd{kdF} $-$ \kd{kdM0} & $+0.033^{\dagger}$ \tiny($+0.029$, $+0.037$) & $+0.030^{\dagger}$ \tiny($+0.023$, $+0.037$) & $-0.064^{\dagger}$ \tiny($-0.071$, $-0.053$) & $-0.009^{\dagger}$ \tiny($-0.016$, $-0.004$) \\
\kd{kdF} $-$ \kd{kdA4} & $+0.030^{\dagger}$ \tiny($+0.028$, $+0.033$) & $+0.030^{\dagger}$ \tiny($+0.026$, $+0.035$) & $-0.069^{\dagger}$ \tiny($-0.081$, $-0.061$) & $-0.011^{\dagger}$ \tiny($-0.015$, $-0.006$) \\
\kd{hardA} $-$ \kd{kdA4} & $+0.023$ \tiny($+0.020$, $+0.026$) & $+0.025$ \tiny($+0.023$, $+0.026$) & $-0.065^{\dagger}$ \tiny($-0.069$, $-0.060$) & $-0.011^{\dagger}$ \tiny($-0.018$, $-0.004$) \\
\bottomrule
\end{tabular}
\end{table*}

\begin{table}[t]
\centering
\caption{Teacher-swap shifts, exploratory. Each row is a difference in differences against \kd{direct}, so zero means no preference for the own teacher. Intervals are 95\% cluster bootstrap intervals over the units shown.}
\label{tab:swaps}
\footnotesize
\setlength{\tabcolsep}{4pt}
\begin{tabular}{@{}lrr@{}}
\toprule
Comparison & Units & Shift (95\% CI) \\
\midrule
\multicolumn{3}{@{}l}{\emph{H1 replication}} \\
\quad \kd{kdA4}: A vs B & 18 & $+0.0071$ \tiny($+0.0046$, $+0.0091$) \\
\quad \kd{kdB4}: B vs A & 18 & $+0.0255$ \tiny($+0.0235$, $+0.0280$) \\
\multicolumn{3}{@{}l}{\emph{single vs single}} \\
\quad \kd{kdM0}: member 0 vs B & 18 & $+0.0171$ \tiny($+0.0150$, $+0.0193$) \\
\multicolumn{3}{@{}l}{\emph{identity only}} \\
\quad \kd{kdM0}: member 0 vs member 1 & 18 & $+0.0224$ \tiny($+0.0207$, $+0.0242$) \\
\quad \kd{kdM1}: member 1 vs member 0 & 18 & $+0.0193$ \tiny($+0.0170$, $+0.0217$) \\
\multicolumn{3}{@{}l}{\emph{anchor}} \\
\quad \kd{hardA}: A vs B & 18 & $-0.0000$ \tiny($-0.0021$, $+0.0019$) \\
\multicolumn{3}{@{}l}{\emph{family swap}} \\
\quad \kd{kdC}: C vs A & 9 & $+0.0432$ \tiny($+0.0297$, $+0.0573$) \\
\bottomrule
\end{tabular}
\end{table}

\begin{table}[t]
\centering
\caption{Detection ageing of the teacher ensemble and of its members (exploratory): slope of unknown-detection AUROC per week since training, one intercept per start date.}
\label{tab:drift}
\begin{tabular}{lrr}
\toprule
Model & Energy & MSP \\
\midrule
Teacher A (ensemble) & $-0.00285$ & $-0.00306$ \\
member mean & $-0.00285$ & $-0.00282$ \\
teacherA\_0 & $-0.00278$ & $-0.00282$ \\
teacherA\_1 & $-0.00306$ & $-0.00291$ \\
teacherA\_2 & $-0.00261$ & $-0.00270$ \\
teacherA\_3 & $-0.00289$ & $-0.00283$ \\
teacherA\_4 & $-0.00291$ & $-0.00287$ \\
\bottomrule
\end{tabular}
\end{table}

\begin{table}[t]
\centering
\caption{Calendar overlap between the three start dates' test windows (Fig.~\ref{fig:design}(a)): 15 of the 18 units share at least one test week with another start date, 55 unit-weeks in total.}
\label{tab:overlap}
\begin{tabular}{rrrrr}
\toprule
Start & Windows & First week & Last week & Shared unit-weeks \\
\midrule
11 & 9 & 16 & 51 & 23 \\
24 & 6 & 29 & 51 & 23 \\
37 & 3 & 42 & 51 & 9 \\
\bottomrule
\end{tabular}
\end{table}

\subsection{A Dedicated Open-Set Detector}
\label{sec:featurescores}

The confirmatory tests score unknown traffic from the logits alone. Under those scores Teacher A and the directly trained student differ by $0.000$ energy AUROC, and Teacher A and the tuned distilled student by $0.001$. H2 and H5 therefore compare students in a regime where the teacher has no detection advantage to pass on, which makes their negative results hard to interpret. A student cannot be shown to miss what its teacher never had.

\begin{figure*}[t]
\centering
\includegraphics[width=\textwidth]{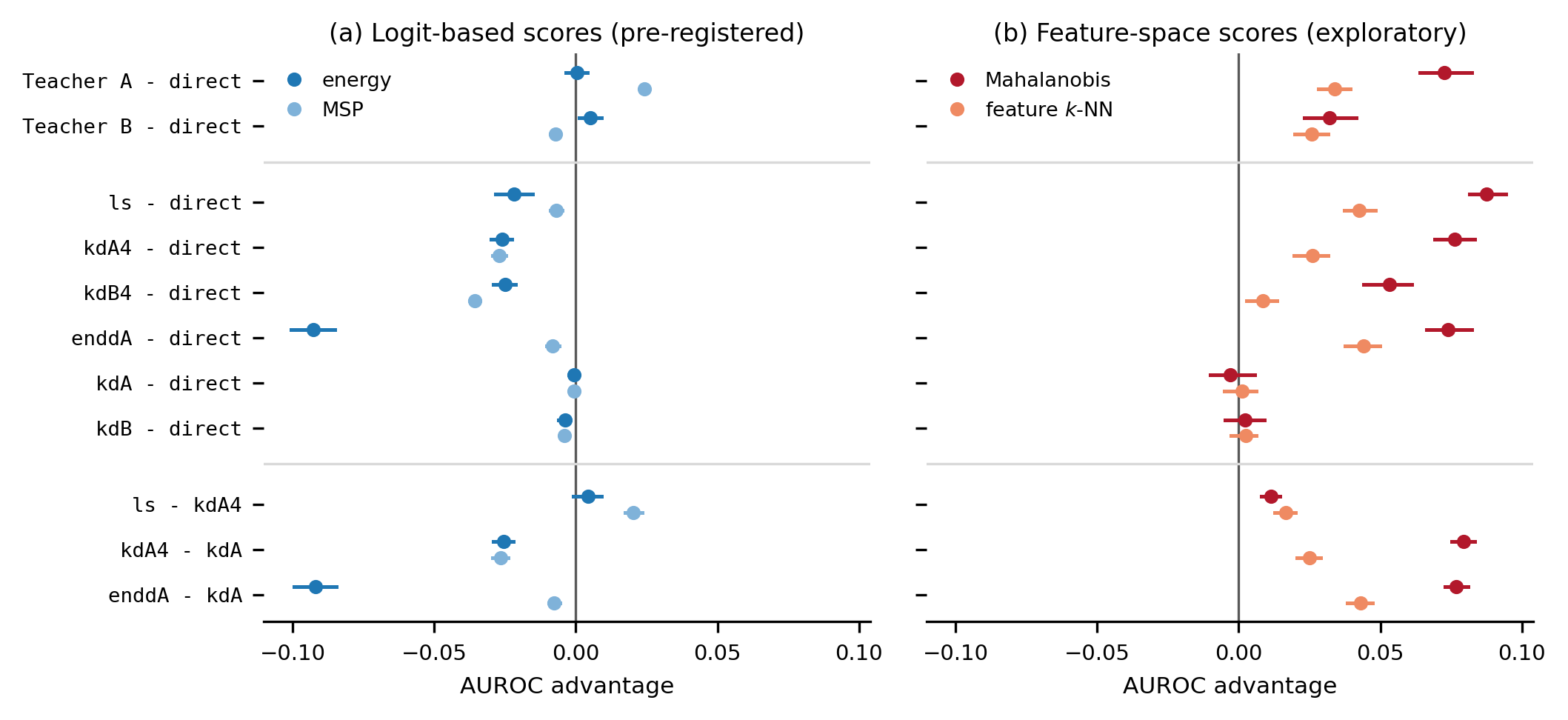}
\caption{Unknown-traffic detection advantage over the directly trained student, by scoring rule (exploratory; 18 test windows): (a) the pre-registered logit-based scores; (b) the feature-space scores. Markers are pooled estimates, bars are 95\% cluster bootstrap intervals, and both panels share one scale. The top block compares models of different feature dimensionality and is indicative; the lower blocks compare identical architectures.}
\label{fig:advantage}
\end{figure*}

The teacher's advantage might be genuinely absent, or it might be real and merely invisible to a logit-based score. To tell these apart we re-scored every model already trained, using two detectors built on its penultimate features: a Mahalanobis distance to the nearest class mean under a tied, shrunk covariance, and the distance to the $10$th nearest training flow. Both are standard open-set detectors, and both read the representation rather than the output layer. No model was retrained; the scores come from checkpoints and evaluation windows that already existed. Fig.~\ref{fig:advantage} reports them beside the pre-registered scores, and Table~\ref{tab:logitscores} reports the same quantity for the exploratory conditions, which the sections that follow discuss one at a time.

The answer is that it was the scoring rule, and specifically the energy score. Pooled over the 18 test windows, Teacher A detects unknown traffic $0.073$ AUROC better than the directly trained student under the Mahalanobis score (95\% CI $0.064$ to $0.082$) and $0.034$ better under the feature $k$-NN score ($0.028$ to $0.039$), against $0.000$ under the energy score ($-0.004$ to $0.004$). A teacher advantage of the size the objection presumes does exist. It is invisible to the energy score, and only partly visible to MSP, which puts it at $0.024$.

What the student does with that advantage depends sharply on the temperature. The conventional-temperature student \kd{kdA4} recovers essentially all of it, $+0.076$ Mahalanobis AUROC over \kd{direct} ($0.069$ to $0.083$) against the $+0.073$ available. The accuracy-optimal student \kd{kdA} recovers none, at $-0.003$ with an interval spanning zero ($-0.010$ to $0.006$). This is the same temperature split that H1 finds with the teacher-swap test, but now in an outcome an operator would act on rather than in a rank correlation. It also sharpens the reading of H2, whose arm is the tuned one: that arm transfers nothing because its targets at $T=1$ are nearly one-hot, not because there was nothing to transfer.

The direction of the effect can even reverse with the scoring rule. \kd{kdA4} is \emph{worse} than \kd{direct} under both logit scores ($-0.026$ energy, $-0.020$ MSP) and better under both feature-space scores. The same model is the better or the worse detector depending only on which part of it is read.

None of this makes the result an argument for distillation. Label smoothing involves no teacher at all, yet it transfers more of the advantage than distillation does. We test this as a paired difference rather than by comparing two intervals, because two overlapping intervals do not establish that their difference excludes zero. \kd{ls} exceeds \kd{kdA4} by $+0.011$ Mahalanobis AUROC ($0.008$ to $0.015$), by $+0.017$ under feature $k$-NN ($0.013$ to $0.020$) and by $+0.012$ under MSP; only under the energy score is the difference indistinguishable from zero ($+0.004$). The pre-registered conclusion, that a cheaper and teacher-free control matches or beats distillation, therefore survives in exactly the regime where a real advantage exists.

The feature-space comparison also loses a control. Temperature scaling rescales logits and leaves the penultimate features untouched, so \kd{directTS} and \kd{direct} are identical under both feature-space scores by construction, and \kd{directTS} is omitted from Fig.~\ref{fig:advantage}.

A second caution applies to the MSP column specifically, and we state it because we found it late. The per-flow score files store MSP in \texttt{float16}, which cannot represent a probability between $0.9995$ and $1$, so every more-confident flow collapses onto exactly $1.0$: 48\% of Teacher B's flows, 40\% of \kd{kdA4}'s, 20\% of \kd{direct}'s and none of \kd{ls}'s, which label smoothing prevents from saturating. The resulting ties deflate AUROC in proportion to a model's confidence, which biases the exploratory MSP column against the teachers and the $T=4$ students and leaves \kd{ls} untouched. The two MSP figures quoted above are therefore taken from the full-precision evaluation that produced Table~\ref{tab:headline}, not from the score files, and the MSP intervals in Table~\ref{tab:logitscores} and Fig.~\ref{fig:advantage} are indicative only. The energy score is unaffected, since its values do not pile up against a representable ceiling: recomputing it at full precision moves every window by less than $0.0001$ AUROC. The confirmatory MSP tests are unaffected in decision terms, because they compare students with similar confidence to one another; recomputed at full precision, H2[MSP] and H5[MSP] keep their sign, their magnitude to within $0.001$ and their outcome. Section~\ref{sec:conclusion} records the storage decision as a limitation.

One further caution applies to reading Fig.~\ref{fig:advantage} across rows. The teachers' features are 600- and 1{,}200-dimensional and the students' are 128-dimensional, and a Mahalanobis distance is not scale-free across dimensionality. Teacher-versus-student rows should be read as indicative. The student-versus-student rows, which carry the argument above, compare models of identical architecture and are not affected.

Finally, the ensemble-distribution-distillation student, the worst detector in the grid under the energy score ($-0.092$ against \kd{kdA}, CI $-0.099$ to $-0.085$), is among the best under both feature-space scores ($+0.077$ and $+0.043$ against \kd{kdA}, both intervals clear of zero). That is consistent with the mechanism proposed in Section~\ref{sec:discussion}, where the proxy-Dirichlet objective caps the total concentration, distorting exactly the logit scale the energy score reads while leaving the representation intact. H5's negative result is therefore about logit-based scores, not about what distribution distillation preserves.

\subsection{Separating Teacher Identity from Ensembling, Width and Family}
\label{sec:swaps}

The registered swap compares an ensemble of five networks against a single wider one of the same family, so a student's shift toward its own teacher could reflect ensembling or width rather than that teacher's identity. The conditions in Table~\ref{tab:swaps} separate them. All run at $T=4$ and all are analysed with the H1 statistic, the difference in differences against the directly trained student, so that they sit on the same scale as the pre-registered estimates of Table~\ref{tab:hypotheses}.

\emph{Identity alone suffices.} \kd{kdM0} and \kd{kdM1} are distilled from two members of Teacher A's ensemble, which share an architecture, a width and a training set and differ only in random seed. Each student follows its own member more than the other, by $+0.022$ ($0.021$ to $0.024$) and $+0.019$ ($0.017$ to $0.022$). Nothing about ensembling or capacity distinguishes these teachers, so whatever the student is inheriting is tied to the particular function its teacher computes. The effect is larger than the registered \kd{kdA4} arm's $+0.007$, so the original swap understated rather than manufactured it.

\emph{None of it comes from label agreement.} \kd{hardA} is trained by cross-entropy on Teacher A's top-1 predictions, so it copies the teacher's decisions and receives no soft-target information. Its shift toward Teacher A is $-0.0000$, with an interval of $-0.002$ to $0.002$ and a one-sided $p$ of $0.57$. Copying a teacher's labels produces no measurable preference for that teacher's per-flow scores. \kd{hardA} is otherwise an ordinary student, matching the directly trained student's closed-set macro-F1 on the validation week ($0.913$ against $0.913$, accuracy $0.941$ against $0.941$), so this is not a degenerate control that failed to learn. It shows instead that the soft targets, and not the decisions, carry the inherited pattern.

\emph{Across model families the effect is larger.} Teacher C is a transformer encoder over the packet sequence (3.5M parameters), trained on start date 11 only and within 0.6 macro-F1 points of Teacher A on its validation week. A student distilled from it follows it over Teacher A by $+0.043$ ($0.030$ to $0.057$), the largest shift measured. Read together, the four conditions fall into the order one would hope a real effect to take, from $0.000$ for label copying, through $0.007$ to $0.026$ between an ensemble and a wide single model and $0.019$ to $0.022$ between two seeds of the same architecture, to $0.043$ across architectures. This last row rests on nine units from one start date and is reported as a probe.

\subsection{A Null Scale for the Teacher Preference}
\label{sec:null}

H1's estimates, $+0.007$ and $+0.026$, raise the question of what counts as a large shift. A useful yardstick is how much a student's preference between two teachers varies when nothing should make it vary at all. We measure this on the \emph{undistilled} student. For each pair of Teacher A's five members we compute $|\rho(\kd{direct}, m_i) - \rho(\kd{direct}, m_j)|$ over all units and seeds. The five members are equally good and come from the same architecture and data, so any difference here is noise. The 540 resulting values have a median of $0.006$, a 95th percentile of $0.016$ and a maximum of $0.025$.

Against that scale the results divide. The identity-only shifts ($+0.019$, $+0.022$), the \kd{kdB4} arm ($+0.026$) and the family swap ($+0.043$) all exceed the 95th percentile of this variation. The \kd{kdA4} arm's $+0.007$ does not; it sits close to the median. H1 is still supported, because that arm's interval excludes zero, but the effect is small relative to the spread one sees between two arbitrary teachers of the same family.

We report this comparison without resolving it, because the two quantities are not strictly commensurable. H1's difference-in-differences already removes the baseline preference that this null measures. A reader asking whether $+0.007$ is a large number nonetheless deserves to see the scale it sits on.

\subsection{Aggregation as an Explanation for the Ageing}
\label{sec:drift}

H3 reversed, in that the teacher--student gap narrows over the year. One explanation, which Section~\ref{sec:discussion} would otherwise have to offer untested, is that this is an artefact of aggregation, since five members' energy scores averaged into one might drift more than any single model's. It is not. Fitting the same fixed-effects slope to each member separately and to the ensemble settles it for the energy score and, we should say plainly, does not settle it for MSP. Under the energy score the ensemble ages at $-0.00285$ AUROC per week against a member mean of $-0.00285$, agreeing to five decimals, with individual members between $-0.00261$ and $-0.00306$ (Table~\ref{tab:drift}). The two feature-space scores agree to within $0.00002$. Aggregation explains none of the energy-score result, which is the one H3 reversed on and the one Fig.~\ref{fig:gap}(a) draws.

Under MSP the answer is weaker. Table~\ref{tab:drift}'s MSP column is computed from the quantised score files described in Section~\ref{sec:featurescores}, and the saturation it induces differs between the ensemble and its members, so we refit that column at full precision. The ensemble then ages at $-0.00316$ per week against its members' $-0.00307$, an excess of $0.00009$, while the MSP gap narrows at $0.00036$ per week. Aggregation accounts for about a quarter of the MSP narrowing rather than all of it, so it is a partial explanation there and not a sufficient one. The mechanism is settled for one co-primary score and partly open for the other.

What remains for the energy score is the other reading, that the teacher and its members age at one rate and the 101k student ages more slowly. That reading now rests on a tested mechanism rather than a guess, and the deployment implication follows from it; under MSP the teacher never loses its lead in any case, so nothing in the recommendation turns on which mechanism operates there.

\subsection{Dependence Between Start Dates}
\label{sec:overlap}

The 18 units come from three start dates whose evaluation windows overlap in calendar time, so they are not independent replicates. Of the 18, 15 contain at least one week that another start date also tests; 55 unit-weeks are shared in total (Fig.~\ref{fig:design}(a) and Table~\ref{tab:overlap}). The cluster bootstrap resamples day $\times$ service clusters, which handles dependence within a window but not between series, so the confirmatory intervals are narrower than fully independent replication would have given.

Fitting H3's trend separately per start date shows that the reversal is not driven by one series. For the \kd{kdA4} arm the slopes are $-0.00044$, $-0.00055$ and $-0.00112$ AUROC per week at start dates 11, 24 and 37, all negative. For the tuned \kd{kdA} arm they are $-0.00050$, $+0.00014$ and $-0.00152$, where the middle series is flat, which is why the pooled estimate is small. No start date shows the widening gap H3 predicted.

\subsection{Student Capacity}
\label{sec:capacity}

The registered grid uses one student width. Repeating \kd{direct} and \kd{kdA4} at widths 16 and 96 on start date 11 gives a capacity gradient (Table~\ref{tab:logitscores}). Against the registered width of 48, the width-16 student is a clearly worse unknown detector ($-0.024$ energy AUROC, CI $-0.033$ to $-0.018$) and the width-96 student is no better ($+0.002$, interval spanning zero). Detection does not improve with student capacity over this range, having already saturated by width 48.

Distillation does not close the gap at any width. Table~\ref{tab:logitscores} measures every row against the registered width-48 \kd{direct} student, so the equal-width comparison is a difference of two of its rows: under the energy score \kd{kdA4} is worse than the directly trained student of its own size by $0.030$ at width 16, $0.026$ at width 48 and $0.017$ at width 96. The penalty does grow as the student shrinks, from $0.017$ at width 96 to $0.030$ at width 16, but distillation trails the equally sized direct student at every width, and the gradient is far shallower than a comparison against the width-48 baseline alone would suggest. The shortcut experiment's conclusion is stated for a single student size in Section~\ref{sec:discussion}; this sweep shows that the relevant property, a small student detecting unknowns at least as well as distillation makes it, holds across a sixfold range of student width.

\subsection{Distilling the Representation Instead of the Logits}
\label{sec:featurekd}

Section~\ref{sec:featurescores} places the teacher's detection advantage in its penultimate representation rather than in its logits, which makes an objective that matches representations the obvious thing to try. We trained one. \kd{kdF} uses similarity-preserving distillation \cite{tung2019}, which matches the batch-wise pairwise similarity structure of the teacher's features instead of its output distribution, from ensemble member 0 at three seeds on start date 11. That is the same teacher \kd{kdM0} distils the logits of, so the two students differ in what they match and in nothing else.

The arm is not degenerate. Its closed-set accuracy on the validation week is $0.941$ against the directly trained student's $0.941$, and its macro-F1 $0.912$ against $0.913$, so the feature penalty leaves the classification objective intact. Nor is the penalty inert: \kd{kdF} ends training at a loss of $0.257$ where \kd{direct} reaches $0.182$.

What it does not do is transfer the teacher's advantage. Over the nine windows start date 11 tests, \kd{kdF} gains $+0.014$ Mahalanobis AUROC over \kd{direct} with an interval spanning zero ($-0.010$ to $+0.033$) and $+0.018$ under feature $k$-NN ($+0.001$ to $+0.028$), against the $+0.073$ the teacher has available. Under both logit scores it is indistinguishable from a student trained without any teacher at all.

Compared directly with logit distillation from the same teacher, on the windows both arms share, it transfers substantially less. \kd{kdF} is $0.064$ Mahalanobis AUROC \emph{below} \kd{kdM0} ($-0.071$ to $-0.053$) and $0.009$ below it under feature $k$-NN ($-0.016$ to $-0.004$); against the registered \kd{kdA4} arm the gap is $0.069$. Matching the teacher's representation transferred less of the teacher's representation-space detector than matching its logits did. What \kd{kdF} avoids is the cost that logit distillation pays elsewhere: it is $+0.033$ better than \kd{kdM0} under the energy score and $+0.030$ under MSP, both intervals clear of zero. Across all four scoring rules it behaves very nearly like the directly trained student, gaining neither the advantage nor the damage.

One explanation is that the objective and the detector read different things. Similarity-preserving distillation matches a row-normalised Gram matrix, which fixes the relative similarity structure of a batch and leaves invariant the rotations and rescalings of the feature space that a Mahalanobis distance to class means is not invariant to. A student can reproduce its teacher's pairwise similarities closely while placing the classes differently in absolute terms, and it is the absolute arrangement that the detector reads.

This arm carries a caveat the others do not, and it is the reason the result is reported as a probe. The loss weight was fixed at $\beta = 100$ and never tuned, because the study's tuning budget was spent before the pre-registration was frozen and was not reopened. A single untuned setting cannot separate ``feature distillation does not transfer this'' from ``$\beta$ was wrong,'' and the training loss shows only that the penalty was active, not that it was near the value that would matter. The conclusion we draw is about what remains to be done rather than about feature distillation: the objective that matches representations is worth testing with a tuned weight and on more than one start date, and Section~\ref{sec:conclusion} states it that way.

\section{Discussion}
\label{sec:discussion}

\subsubsection*{What transfers, and what does not}
The teacher-swap control answers the study's question with some precision. Distillation at a conventional temperature does hand the student something specific to its teacher. The per-flow pattern of unknown-scores shifts measurably toward the teacher that produced the soft targets, and away from an equally accurate teacher that did not (H1). But that pattern is all that transfers. The teacher's \emph{quality} as an unknown detector, its calibration and its robustness to shortcuts do not arrive with it. Against the cheapest possible control, the same student trained directly and temperature-scaled, distillation buys nothing on any unknown-detection outcome, and against label smoothing it trades detection for calibration. The inheritance is indiscriminate: along with the score pattern, the $T=4$ student takes on the teacher's poorer detection of near unknowns.

The temperature result reconciles two findings that would otherwise conflict. Tuned for accuracy, distillation chooses $T=1$, at which a 96\%-accurate teacher's targets are nearly one-hot and nothing teacher-specific survives; the student is then a directly trained student by another name. At $T=4$ the teacher's signal comes through, at a cost of 1.6 macro-F1 points for Teacher A's arm and 1.7 for Teacher B's. Practitioners who tune distillation for accuracy, which is the norm in the traffic-classification literature we surveyed, are therefore not transferring the properties that distillation is often assumed to preserve, and practitioners who keep the textbook temperature are paying accuracy for a pattern that does not improve detection.

\subsubsection*{The gap closes instead of opening}
We expected the compressed student to age faster than its teacher. The opposite happened. Over 35 weeks of real traffic the 11.3M-parameter ensemble lost unknown-detection AUROC faster than the 101k student, and in two of the three replicates the student was the better detector from about the third month. One reading of that is an artefact: the ensemble's energy score aggregates five members' logit scales, so it might drift more than any single model's. For the energy score this is wrong. Section~\ref{sec:drift} fits the ageing slope to each member separately and finds the ensemble ageing at exactly its members' rate ($-0.00285$ per week against a member mean of $-0.00285$). Under MSP the same test explains about a quarter of the narrowing, so aggregation contributes there without accounting for it. The alternative reading, for the score on which the reversal occurs, is that a higher-capacity model fits more of the training period's transient structure, which then decays, so that the teacher ages at its intrinsic rate while the 101k student ages more slowly. Either way the deployment implication is the same, and it is the opposite of the usual worry. In this setting the small model is not the fragile one. The exception is label smoothing, whose gap to the teacher widened, so the effect is not a property of small models in general but of how they are trained. Both teacher and students lose about 20 macro-F1 points over the year. That is the same order as the decline this capture is known to induce within months of training~\cite{luxemburk2024year22}, and it is the degradation that motivates the recent critique of static traffic-classification benchmarks~\cite{ntccrisis}. The drift we measure is a property of the traffic, not an artefact of our small student or our service split. Nothing in this study mitigates drift; it only shows that compression does not make it worse.

\subsubsection*{Shortcuts follow capacity, not training}
Whether a student relies on a shortcut it can see is governed by its size. At every shortcut reliability below 1, both 101k students leaned on the planted feature about twice as hard as the 2.3M teacher, regardless of whether they were distilled; at $\rho=1$ every model collapses and the ratio falls to about 1.1. This agrees with prior findings that compressed models bear a disproportionate share of a model family's brittleness~\cite{hooker2020}, and it means that distillation is not a route to the teacher's shortcut robustness. What distillation \emph{does} carry is confidence. A teacher that learned to lean on a shortcut produces, through its soft targets alone, a student that is more over-confident on data where the shortcut does not exist. That the same student detected unknowns slightly better is a reminder that calibration and open-set ranking are different quantities, and that improving one is no evidence about the other.

\subsubsection*{Where we disagree with the literature}
\emph{Label smoothing.} Several studies report that label smoothing degrades out-of-distribution detection by collapsing the geometry of the penultimate layer~\cite{softlabelsood,lsembedding,adaptivels,lsselective}. Our grid does not reproduce that. Under the two logit scores the label-smoothed student is ahead of every distilled student at $T=4$ and of EnDD, though behind the undistilled student and the $T=1$ arms. Under the feature-space scores it is ahead of the teacher: best in the grid under the Mahalanobis score, and under feature $k$-NN within 0.001 of EnDD at the top, a margin we do not read as a ranking (Section~\ref{sec:featurescores}). A student whose penultimate geometry had collapsed would not be the easiest of all to fit a class-conditional density to. We do not think the prior findings are wrong; we think the regime differs. Those studies use tens of classes, image features and a smoothing strength of 0.1. Ours has 102 fine-grained services with heavy overlap, flow-level features, a tuned strength of 0.05, and unknowns that are themselves TLS services rather than a different modality. The degradation reported for label smoothing may depend on the OOD data being far from the training distribution, which near-unknown services are not. Reproducing the geometric analysis of those papers on this dataset would settle the point and is left for future work.

\emph{Distribution distillation.} Ensemble distribution distillation and its descendants report students that match or exceed their teacher ensembles at uncertainty estimation~\cite{logitendd,selfdistdist,credalendd}. Our EnDD student is the worst unknown detector in the grid with the energy score. We used the proxy-Dirichlet objective rather than maximum-likelihood EnDD, because the latter was unstable with 102 classes in our pilot, and the proxy target caps the precision at $10^4$, which leaves the loss dominated by matching the total concentration, that is the logit scale, instead of the members' disagreement. Our teacher ensemble's members also disagree little (their mean energy AUROC is within 0.015 of the ensemble's), so there is little epistemic signal to distil. A negative result under these conditions should be read as ``proxy EnDD did not preserve detection for a low-diversity ensemble on a 100-class task'', not as a refutation of distribution distillation.

\subsubsection*{Limitations}
Novelty is simulated. No service first appears mid-year in CESNET-TLS-Year22, so unknown services are held out, and the near/far split is by the dataset's own category labels. The study is one dataset, one network, one year. The confirmatory grid uses a single teacher architecture at two widths; the transformer teacher of Section~\ref{sec:swaps} is exploratory and runs on one start date, so the family swap is a probe rather than a replication. The confirmatory results concern \emph{logit} distillation, that is soft targets at the output layer. The exploratory feature-distillation arm (similarity-preserving distillation from one ensemble member) carries an untuned loss weight, so a null result from it would be weak evidence of absence. Feature-based distillation remains the most important question this study leaves open, all the more so now that Section~\ref{sec:featurescores} places the teacher's advantage in the representation and not in the logits. The shortcut experiment plants a synthetic feature, runs on the validation week, and uses a single student width, so it supports the claim that both 101k-parameter students rely about twice as much on the planted feature as the 2.3M-parameter teacher, not that reliance varies smoothly with capacity. Natural shortcut fields (server IP, ASN, JA3) were excluded from the inputs, not tested. H1's support rests more on \kd{kdB4} than on \kd{kdA4}, and the latter's null at start date 11 is unexplained. Baselines and the deployment cost were measured on the validation week and a laptop CPU. The per-flow score files store the maximum softmax probability in \texttt{float16}, whose spacing below $1$ is coarse enough that confident flows saturate; the exploratory MSP estimates computed from those files are biased against confident models, and the figures we quote for MSP are taken from the full-precision evaluation instead (Section~\ref{sec:featurescores}). The energy and feature-space scores are unaffected, and no confirmatory outcome changes, but the score files as deposited carry the limitation. The pre-registration was frozen on validation data that had already informed eight design decisions, all documented; the test windows and their unknown services were not touched until the freeze, and the public registration certifies the same document one day later.

\section{Conclusion and Future Work}
\label{sec:conclusion}

A logit-distilled traffic classifier inherits its teacher's habits, and, at the conventional temperature and only where a scoring rule can see it, an ability that a teacher was never needed to obtain. The per-flow score pattern transfers, it transfers from the specific teacher used, and it carries that teacher's over-confidence and its blind spots along with it. What a network operator would want from distillation is unknown-traffic detection that holds up, calibration that can be trusted and robustness to shortcuts. Under the logit-based scores the study registered, and against the accuracy-tuned arm, a directly trained small model already has all three to the same degree, and keeps them at least as well over a year of drift. Where a feature-space detector does see a teacher advantage, the conventional-temperature student inherits it, but so does label smoothing, which needs no teacher. If the goal is a small model that detects unknown traffic, this study finds no reason to train a large one first and distil from its outputs.

The obvious next step is feature-based distillation. Since the teacher's advantage lives in the representation and not in the logits, an objective that matches representations is the natural thing to try, and our one attempt at it (Section~\ref{sec:featurekd}) transferred less of that advantage than logit distillation did, not more. But that attempt used a single untuned loss weight on one start date, which is too thin to conclude from. A tuned weight, several start dates, and objectives that constrain the absolute arrangement of the representation and not only its pairwise similarities would settle it. Teacher scale is the second gap: our teachers are 2.3M-parameter networks, the regime in which a swap control is affordable, and whether a hundred-million-parameter traffic foundation model transfers more, or transfers something different, needs the same protocol at a scale we could not reach. The ageing result, finally, is measured but not explained. We show that the ensemble ages at its members' rate and that the small student ages more slowly, without knowing what drives either. Relating ageing to what each model fits in the training period's transient structure would turn the measurement into an explanation.

\section*{Reproducibility}
Code, the frozen service split, the per-window metric tables and the analysis are released at \url{https://github.com/Mahmoud-Abbasi-svg/kd-encrypted-traffic-inheritance}, with the commit that froze the pre-registration tagged \texttt{prereg-frozen} so that the registered protocol and the analysis code implementing it can be checked against each other. The pre-registration itself is at OSF \url{https://osf.io/rts6n}. The per-flow unknown-scores for every model and evaluation window (129 files, 8.2\,GB) and the 206 model checkpoints exceed what a source repository can hold, so they are deposited at Zenodo under CC BY 4.0 (\url{https://doi.org/10.5281/zenodo.22916038}), with a manifest listing every file and its digest. The deposit contains derived scores and trained weights only, not the captured flows.

\balance
\bibliographystyle{IEEEtran}
\bibliography{refs}

\begin{thebibliography}{10}
\providecommand{\url}[1]{#1}
\csname url@samestyle\endcsname
\providecommand{\newblock}{\relax}
\providecommand{\bibinfo}[2]{#2}
\providecommand{\BIBentrySTDinterwordspacing}{\spaceskip=0pt\relax}
\providecommand{\BIBentryALTinterwordstretchfactor}{4}
\providecommand{\BIBentryALTinterwordspacing}{\spaceskip=\fontdimen2\font plus
\BIBentryALTinterwordstretchfactor\fontdimen3\font minus \fontdimen4\font\relax}
\providecommand{\BIBforeignlanguage}[2]{{%
\expandafter\ifx\csname l@#1\endcsname\relax
\typeout{** WARNING: IEEEtran.bst: No hyphenation pattern has been}%
\typeout{** loaded for the language `#1'. Using the pattern for}%
\typeout{** the default language instead.}%
\else
\language=\csname l@#1\endcsname
\fi
#2}}
\providecommand{\BIBdecl}{\relax}
\BIBdecl

\bibitem{luxemburk2024year22}
K.~Hynek, J.~Luxemburk, J.~Pe{\v{s}}ek, T.~{\v{C}}ejka, and P.~{\v{S}}i{\v{s}}ka, ``{CESNET-TLS-Year22}: A year-spanning {TLS} network traffic dataset from backbone lines,'' \emph{Scientific Data}, vol.~11, no.~1, p. 1156, 2024.

\bibitem{etbert}
X.~Lin, G.~Xiong, G.~Gou, Z.~Li, J.~Shi, and J.~Yu, ``{ET-BERT}: A contextualized datagram representation with pre-training transformers for encrypted traffic classification,'' in \emph{Proc. ACM Web Conference}, 2022.

\bibitem{yatc}
R.~Zhao, M.~Zhan, X.~Deng, Y.~Wang, Y.~Wang, G.~Gui, and Z.~Xue, ``Yet another traffic classifier: A masked autoencoder based traffic transformer with multi-level flow representation,'' in \emph{Proc. AAAI}, 2023.

\bibitem{netfound}
S.~Guthula, R.~Beltiukov, N.~Battula, W.~Guo, and A.~Gupta, ``{netFound}: Foundation model for network security,'' \emph{arXiv preprint arXiv:2310.17025}, 2023.

\bibitem{hinton2015}
G.~Hinton, O.~Vinyals, and J.~Dean, ``Distilling the knowledge in a neural network,'' \emph{arXiv preprint arXiv:1503.02531}, 2015.

\bibitem{netclus}
Z.~Huang, C.~Lin, W.~Zhang, X.~Meng, and Y.~Zhang, ``{Distillation-Enhanced Clustering Acceleration for Encrypted Traffic Classification},'' \emph{arXiv preprint arXiv:2508.02282}, 2025.

\bibitem{merlot}
Y.~Chen, R.~Li, Z.~Zhao, and H.~Zhang, ``{MERLOT: A Distilled LLM-based Mixture-of-Experts Framework for Scalable Encrypted Traffic Classification},'' \emph{arXiv preprint arXiv:2411.13004}, 2024.

\bibitem{resaware}
C.~Fan, W.~Wang, W.~Huang, Z.~Ding, J.~Shi, L.~Cui, Z.~Hao, and X.~Yun, ``{ResAware: Cross-Environment Website Fingerprinting via Resource-Privileged Distillation},'' \emph{arXiv preprint arXiv:2606.17462}, 2026.

\bibitem{ciphersight}
R.~Song, Q.~Liu, C.~Pan, Z.~Ding, Y.~Xian, C.~Fan, L.~Cui, W.~Wang, and Z.~Hao, ``{CipherSight: Robust Website Fingerprinting via Record-Resource Semantic Supervision under Distribution Shifts},'' \emph{arXiv preprint arXiv:2608.13905}, 2026.

\bibitem{tnsm2026fedkd}
H.~A. Tran and N.-T. Hoang, ``{Towards Efficient and Adaptive Traffic Classification: A Knowledge Distillation-Based Personalized Federated Learning Framework},'' \emph{IEEE Transactions on Network and Service Management}, vol.~23, pp. 594--604, 2026.

\bibitem{ntccrisis}
K.~Jer{\'a}bek, J.~Luxemburk, R.~Pln{\'y}, J.~Koumar, J.~Pe{\v{s}}ek, and K.~Hynek, ``When simple model just works: Is network traffic classification in crisis?'' \emph{arXiv preprint arXiv:2506.08655}, 2025.

\bibitem{luxemburk2022reject}
J.~Luxemburk and T.~{\v{C}}ejka, ``Fine-grained {TLS} services classification with reject option,'' \emph{Computer Networks}, vol. 220, p. 109467, 2023, arXiv:2202.11984.

\bibitem{biasseeker}
C.~Wang, X.~Xie, T.~Wang, and Y.~Cui, ``{Bias in the Shadows: Explore Shortcuts in Encrypted Network Traffic Classification},'' \emph{arXiv preprint arXiv:2601.10180}, 2026.

\bibitem{yuan2020revisiting}
L.~Yuan, F.~E.~H. Tay, G.~Li, T.~Wang, and J.~Feng, ``Revisiting knowledge distillation via label smoothing regularization,'' in \emph{Proc. IEEE/CVF CVPR}, 2020.

\bibitem{functionalkd}
I.~Mason-Williams, G.~Mason-Williams, and H.~Yannakoudakis, ``A functional perspective on knowledge distillation in neural networks,'' \emph{arXiv preprint arXiv:2510.12615}, 2025.

\bibitem{prunedtrees}
Y.~Luo, J.~Tao, X.~Xu, L.~Yu, and K.~Li, ``{Pruned Traffic Trees: Native Semantic Compression with a Protocol-Structured Model Family for Encrypted Traffic Classification},'' \emph{arXiv preprint arXiv:2608.21874}, 2026.

\bibitem{futureinternet2026}
Z.~Li and Y.~Feng, ``{A Lightweight Multi-Classification Model for Identifying Network Application Traffic Using Knowledge Distillation},'' \emph{Future Internet}, vol.~18, no.~4, p. 197, 2026.

\bibitem{mmaeflowmix}
X.~Liu, X.~Fu, F.~Huang, and L.~Zhang, ``{Mean Masked Autoencoder with Flow-Mixing for Encrypted Traffic Classification},'' \emph{arXiv preprint arXiv:2603.29537}, 2026.

\bibitem{stanton2021}
S.~Stanton, P.~Izmailov, P.~Kirichenko, A.~A. Alemi, and A.~G. Wilson, ``Does knowledge distillation really work?'' in \emph{Advances in Neural Information Processing Systems}, 2021, arXiv:2106.05945.

\bibitem{ojha2023}
U.~Ojha, Y.~Li, A.~Sundara~Rajan, Y.~Liang, and Y.~J. Lee, ``What knowledge gets distilled in knowledge distillation?'' in \emph{Advances in Neural Information Processing Systems}, 2023, arXiv:2205.16004.

\bibitem{teacherspet}
M.~Lukasik, S.~Bhojanapalli, A.~K. Menon, and S.~Kumar, ``Teacher's pet: Understanding and mitigating biases in distillation,'' \emph{Transactions on Machine Learning Research}, 2022.

\bibitem{beyonddarkknowledge}
J.~Medina, P.~Honeine, A.~Bensrhair, and A.~Hadachi, ``{Beyond Dark Knowledge: Mixup-Based Distillation for Reliable Predictions},'' \emph{arXiv preprint arXiv:2606.12171}, 2026.

\bibitem{cud2026}
J.~Kim, S.~Kim, R.~Xuan, and H.~Cho, ``{Trust the Uncertain Teacher: Distilling Dark Knowledge via Calibrated Uncertainty},'' \emph{arXiv preprint arXiv:2602.12687}, 2026.

\bibitem{teachercalibration2025}
S.~Kim, S.~Park, J.~Lee, and N.~Kwak, ``{The Role of Teacher Calibration in Knowledge Distillation},'' \emph{arXiv preprint arXiv:2508.20224}, 2025.

\bibitem{kdloses2026}
W.~Wang, ``{Knowledge Distillation Must Account for What It Loses},'' \emph{arXiv preprint arXiv:2604.25110}, 2026.

\bibitem{whotaughtyou}
S.~Wadhwa, C.~Shaib, S.~Amir, and B.~C. Wallace, ``Who taught you that? tracing teachers in model distillation,'' in \emph{Findings of the Association for Computational Linguistics: ACL 2025}, 2025, arXiv:2502.06659.

\bibitem{kddetection2025}
Q.~Shi, A.~Y. Zheng, Q.~Song, and R.~A. Yeh, ``{Knowledge Distillation Detection for Open-weights Models},'' in \emph{Advances in Neural Information Processing Systems}, 2025, arXiv:2510.02302.

\bibitem{antidistill}
Y.~E. Xu, J.~Kirchenbauer, Y.~Savani, A.~Trockman, A.~Robey, T.~Goldstein, F.~Fang, and J.~Z. Kolter, ``{Antidistillation Fingerprinting},'' \emph{arXiv preprint arXiv:2602.03812}, 2026.

\bibitem{tokenprint}
Y.~Wu, S.~Zhao, and J.~Chen, ``{TokenPrint: A Calibrated Token-Space Fingerprint for Language-Model Provenance},'' \emph{arXiv preprint arXiv:2608.08139}, 2026.

\bibitem{darknetopenworld}
J.~Saleem, R.~Islam, and M.~Z. Islam, ``{Open-World Darknet Traffic Recognition Under Leave-One-Service-Out Evaluation},'' \emph{arXiv preprint arXiv:2608.04167}, 2026.

\bibitem{soklabels}
S.~Huang and S.~Yang, ``{SoK: Where Do Flow Labels Come From? Auditing Label Provenance in Encrypted Traffic Benchmarks},'' \emph{arXiv preprint arXiv:2609.02140}, 2026.

\bibitem{unialign}
T.~Wang, X.~Xie, W.~Wang, C.~Wang, and Y.~Cui, ``{UniAlign: A Model-Agnostic Framework for Robust Network Traffic Classification under Distribution Shifts},'' \emph{arXiv preprint arXiv:2605.17575}, 2026.

\bibitem{colossus}
B.~Li, L.~Meng, R.~Song, C.~Pan, T.~Pu, Z.~Ma, Y.~Jiang, L.~Cui, and Z.~Hao, ``{The Colossus with Feet of Clay: Debunking Encrypted Traffic Classifiers under PQC Evolution},'' \emph{arXiv preprint arXiv:2608.22683}, 2026.

\bibitem{tddm}
Z.~Chen, Q.~Yu, Z.~Song, G.~Yang, and W.~Yan, ``{TDDM-Melatt: A Decoupled Memory and Diffusion Framework for Generalizable Encrypted Traffic Classification},'' \emph{arXiv preprint arXiv:2608.30745}, 2026.

\bibitem{asd2026}
S.~M. Raza, O.~Tariq, and J.~Son, ``{Anti-Shortcut Distillation via Temporal Negative Knowledge Transfer},'' \emph{arXiv preprint arXiv:2608.11789}, 2026.

\bibitem{saopd2026}
Y.~Jiang, Y.~Ye, Z.~Tao, X.~Zhuang, Q.~Zhang, H.~Chen, and T.~Li, ``{When Teachers Mislead: Spurious-Signal-Aware On-Policy Distillation},'' \emph{arXiv preprint arXiv:2608.03632}, 2026.

\bibitem{iga2026}
Z.~Cheng, W.~Dai, and J.~Sun, ``{Invariant Gradient Alignment for Robust Reasoning Distillation},'' \emph{arXiv preprint arXiv:2606.05025}, 2026.

\bibitem{medshortcutkd}
C.~Boland, S.~Tsaftaris, and S.~Dahdouh, ``{Preventing Shortcut Learning in Medical Image Analysis through Intermediate Layer Knowledge Distillation from Specialist Teachers},'' \emph{arXiv preprint arXiv:2511.17421}, 2025.

\bibitem{hooker2020}
S.~Hooker, N.~Moorosi, G.~Clark, S.~Bengio, and E.~Denton, ``Characterising bias in compressed models,'' \emph{arXiv preprint arXiv:2010.03058}, 2020.

\bibitem{mtforget}
A.~Mohammadshahi, V.~Nikoulina, A.~Berard, C.~Brun, J.~Henderson, and L.~Besacier, ``{What Do Compressed Multilingual Machine Translation Models Forget?}'' \emph{arXiv preprint arXiv:2205.10828}, 2022.

\bibitem{robustnessdistill}
M.~Du, S.~Mukherjee, Y.~Cheng, M.~Shokouhi, X.~Hu, and A.~H. Awadallah, ``{Robustness Challenges in Model Distillation and Pruning for Natural Language Understanding},'' \emph{arXiv preprint arXiv:2110.08419}, 2021.

\bibitem{malinin2020endd}
A.~Malinin, B.~Mlodozeniec, and M.~Gales, ``Ensemble distribution distillation,'' in \emph{Proc. ICLR}, 2020.

\bibitem{logitendd}
Y.~Fathullah, G.~Xia, and M.~Gales, ``{Logit-Based Ensemble Distribution Distillation for Robust Autoregressive Sequence Uncertainties},'' \emph{arXiv preprint arXiv:2305.10384}, 2023.

\bibitem{selfdistdist}
Y.~Fathullah and M.~J.~F. Gales, ``{Self-Distribution Distillation: Efficient Uncertainty Estimation},'' \emph{arXiv preprint arXiv:2203.08295}, 2022.

\bibitem{credalendd}
K.~Wang, F.~Cuzzolin, D.~Moens, and H.~Hallez, ``{Credal Ensemble Distillation for Uncertainty Quantification},'' \emph{arXiv preprint arXiv:2511.13766}, 2025.

\bibitem{ryabinin2021}
M.~Ryabinin, A.~Malinin, and M.~Gales, ``Scaling ensemble distribution distillation to many classes with proxy targets,'' in \emph{Advances in Neural Information Processing Systems}, 2021.

\bibitem{softlabelsood}
D.~Lee and Y.~Cheon, ``{Soft Labeling Affects Out-of-Distribution Detection of Deep Neural Networks},'' \emph{arXiv preprint arXiv:2007.03212}, 2020.

\bibitem{lsembedding}
D.~Bahri, H.~Jiang, Y.~Tay, and D.~Metzler, ``{Label Smoothed Embedding Hypothesis for Out-of-Distribution Detection},'' \emph{arXiv preprint arXiv:2102.05131}, 2021.

\bibitem{adaptivels}
M.~Xu, J.~Lee, S.~Yoon, and D.~S. Park, ``{Adaptive Label Smoothing for Out-of-Distribution Detection},'' \emph{arXiv preprint arXiv:2410.06134}, 2024.

\bibitem{lsselective}
G.~Xia, O.~Laurent, G.~Franchi, and C.-S. Bouganis, ``{Towards Understanding Why Label Smoothing Degrades Selective Classification and How to Fix It},'' \emph{arXiv preprint arXiv:2403.14715}, 2024.

\bibitem{tung2019}
F.~Tung and G.~Mori, ``{Similarity-Preserving Knowledge Distillation},'' in \emph{Proc. IEEE/CVF International Conference on Computer Vision (ICCV)}, 2019, pp. 1365--1374.

\end{thebibliography}

\begin{IEEEbiographynophoto}{Mahmoud Abbasi}
received the M.Sc. degree in computer software engineering from Islamic Azad University, Mashhad, Iran, in 2017, and the Ph.D. degree in computer engineering from the University of Salamanca, Salamanca, Spain, in 2026, where the doctoral work was supported by a Marie Sk\l{}odowska-Curie Fellowship within the EU Horizon 2020 IoTalentum network (grant no.\ 953442).

Since 2026, Dr.\ Abbasi has been a Researcher with the AIR Institute, Salamanca, Spain. Research interests include machine learning for network traffic monitoring and analysis, intent-based networking, and secure, scalable Internet-of-Things systems. Dr.\ Abbasi has been a Member of IEEE since 2019.
\end{IEEEbiographynophoto}

\end{document}